\documentclass{article}
\usepackage{epsfig}
\usepackage[onecolumn]{emulateapj}

\newcommand{\hepph}[1]{{\tt hep-ph/#1}}

\newcommand{\href}[2]{#1}

\def\lsim{\mathrel{\rlap{\lower4pt\hbox{\hskip1pt$\sim$}}
    \raise1pt\hbox{$<$}}}         
\def\gsim{\mathrel{\rlap{\lower4pt\hbox{\hskip1pt$\sim$}}
    \raise1pt\hbox{$>$}}}         
\def\esim{\mathrel{\rlap{\raise2pt\hbox{$\sim$}}
    \lower1pt\hbox{$-$}}}         

\begin{document}

\centerline{\bf \Large Cosmic antiprotons as a probe }
\smallskip
\centerline{\bf \Large for supersymmetric dark matter?}

\bigskip
\bigskip

\centerline{\large Lars Bergstr{\"o}m\footnote{E-mail address:
lbe@physto.se},
Joakim Edsj{\"o}\footnote{E-mail address:
edsjo@physto.se},
Piero Ullio\footnote{E-mail address:
piero@physto.se}}
\smallskip
\centerline{\em Department of Physics, Stockholm University, Box 6730,
SE-113~85~Stockholm, Sweden}

\submitted{February 1, 1999}

\begin{abstract}
The flux of cosmic ray antiprotons from neutralino annihilations in
the galactic halo is computed for a large sample of models in the MSSM
(the Minimal Supersymmetric extension of the Standard Model).  We also
revisit the problem of estimating the background of low-energy cosmic
ray induced secondary antiprotons, taking into account their
subsequent interactions (and energy loss) and the presence of nuclei
in the interstellar matter.

We consider a two-zone diffusion model, with and without a galactic
wind.  We find that, given the uncertainties in the background
predictions, there is no need for a primary (exotic) component to
explain present data.  However, allowing for a signal by playing with
the uncertainties in the background estimate, we discuss the
characteristic features of the supersymmetric models which give a
satisfactory description of the data.  We point out that in some cases
the optimal kinetic energy to search for a signal from supersymmetric
dark matter is above several GeV, rather than the traditional sub-GeV
region.

The large astrophysical uncertainties involved do not, one the other 
hand, allow the exclusion of any of the MSSM models we consider, on 
the basis of data.

We present besides numerical results also convenient parameterizations
of the antiproton yields of all `basic' two-body final states. We also
give examples of the yield and differential energy spectrum for a set
of supersymmetric models with high rates.

We also remark that it is difficult to put a limit on the
antiproton lifetime from present measurements, since the injection of
antiprotons from neutralino annihilation can compensate the loss from
decay. 

\end{abstract}


\section{Introduction}

The mystery of the dark matter in the Universe remains unsolved.
Among the most plausible candidates are Weakly Interacting Massive
Particles (WIMPs), of which the supersymmetric neutralino is a
favourite candidate from the point of view of particle physics. The
neutralino arises naturally in supersymmetric extensions of the
standard model, and has the attractive feature of giving a relic
density which in large regions of parameter space is adequate to
explain cosmological dark matter.  We will in this paper consider the
neutralino as a dark matter candidate within the Minimal
Supersymmetric extension of the Standard Model (MSSM). For a thorough
review of neutralino dark matter, see \cite{jkg}.

Neutralino dark matter can be and is searched for in several ways:
directly through detection of nuclear recoils and/or ionization
in direct detection
experiments, and indirectly through searches for their annihilation
products from annihilation in the Earth or Sun (for neutrinos)
and the galactic halo.
In this paper, we discuss the detection prospects of antiprotons
from neutralino annihilation in the galactic halo.

As antimatter seems not to exist in large quantities in the
observable Universe, including our own Galaxy, any contribution to
the cosmic ray generated antimatter flux (besides antiprotons also
positrons) from exotic sources may in principle be a good
signature for such sources. Since neutralinos are constrained by
supersymmetry to be Majorana fermions they are their own
antiparticles and therefore the final state in their annihilations
in the halo will contain equal amounts of matter and antimatter
(given the particle physics constraints on CP violating couplings
and in particular on baryon number violation). The excess of
particles would drown in the background of particles from
astrophysical sources, but there is a chance that antiparticles
from this new primary source could be detectable. This issue has
recently come into new focus thanks to upcoming space experiments like
{\sc Pamela} (\cite{pamela}) and {\sc Ams} (\cite{ams}) with
increased sensitivity to the cosmic antimatter flux.

Cosmic ray induced secondary antiprotons are generated mainly through 
$pp\to \bar p + X$ collisions of cosmic ray protons with interstellar 
matter.  For kinematical reasons they are born with a non-zero 
momentum.  The strategy to search for exotic signals has thus been to 
investigate the low-energy region since, e.g., a neutralino-induced 
component does not drop as fast at low energies.  However, as we will 
see, this ideal picture is blurred to a large extent by a `tertiary' 
component caused by scattering with energy loss of the secondary 
antiprotons.  Also, heavier nuclei in the interstellar medium target 
(primarily helium) cause a significant antiproton flux at low energy.  
In particular, it is known that in proton-nucleus collisions 
antiprotons may be produced well below the nominal $pp$ energy 
threshold.  In addition, low energy particles have difficulties 
entering the heliosphere which makes the connection of the measured 
fluxes to the insterstellar ones dependent on a not completely known 
correction due to this solar modulation (which follows the 11-year 
solar cycle).

\begin{figure}
 \centerline{\epsfig{figure=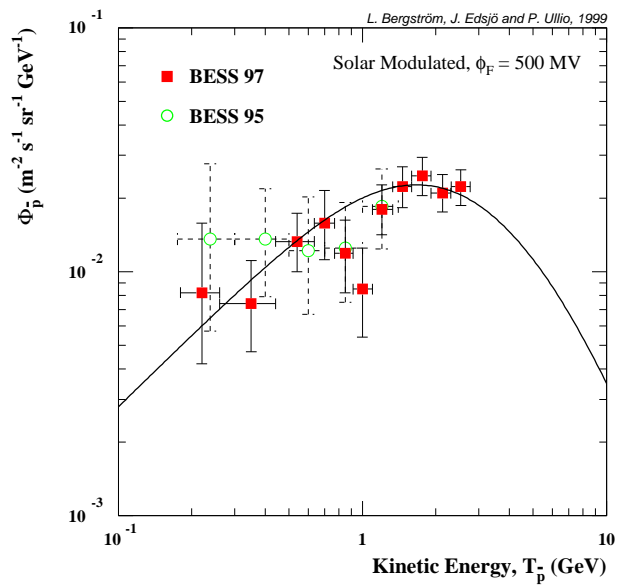,width=0.49\textwidth}}
 \caption[]{The background of antiprotons solar modulated with
 $\phi_F=500$ MV. The {\sc Bess} 95 and 97 data are also shown 
 (\protect\cite{bess95}; \protect\cite{bess97}).}
 \label{fig:back}
\end{figure}

Using reasonable parameters for the computation of all these effects,
we will show that we are able to explain in a satisfactory way the
present experimental data on cosmic ray antiprotons without the need
of a primary component.  This is seen in Fig.~\ref{fig:back}, where
our computed reference distribution for the background is compared to
the recent data from the {\sc Bess} experiment ({\sc Bess} 95:
\cite{bess95}; {\sc Bess} 97: \cite{bess97}). The satisfactory agreement
can be compared to the conclusions of a recent analysis
(\cite{bottinolast}), where the need for an exotic component was more
apparent. The main cause of this difference lies in our improved
treatment of energy loss during propagation and our inclusion of the
helium component of the interstellar medium.  From
Fig.~\ref{fig:back} one can also see that the more recent {\sc Bess}
data (\cite{bess97}) indicate a lower flux at low energy than earlier
data.  If reacceleration effects were included, the need for an exotic
component would be even less.

It is, however, also evident from Fig.\,\ref{fig:back} that the 
statistical sample of antiprotons presently available is very limited, 
so that a new primary component can not yet be ruled out with high 
significance even if the propagation parameters were known.  By 
varying all parameters in this complex astrophysical problem, the room 
for an exotic contribution can, as we will see, be made quite large, 
so it is certainly worthwhile to investigate what a favoured dark 
matter candidate like the MSSM neutralino may yield in terms of a 
signal.  It is apparent from Fig.\,\ref{fig:back} that the low-energy 
tail of cosmic ray induced antiprotons does not fall very rapidly with 
decreasing kinetic energy, so this particular region of phase-space 
may not yield as a nice signature for a dark matter neutralino as 
previously thought.  Therefore, we will also discuss the cases (mainly 
for high-mass neutralinos) when the optimal kinetic energy for finding 
an exotic contribution is above several GeV. Unfortunately, it will 
turn out that the rates in that energy region are not large enough to 
cause a spectral distortion, unless the background of secondary 
antiprotons is considerably smaller and/or the signal is considerably 
larger than our canonical results show.  This lack of spectral 
features in the neutralino-induced antiproton flux causes severe 
fundamental limitations for this indirect method of detecting 
supersymmetric dark matter.

The idea of a dark matter-induced component in the cosmic antiproton
spectrum has a long history. An early report (\cite{Buffington}) of
anomalous excess of cosmic ray antiprotons at low energies led to the
suggestion that annihilation of relic neutralinos could be the source.
Calculations of fluxes have since then been performed with different
degrees of sophistication, ranging from order of magnitude estimates
(\cite{Silk-Sred}), analytical expressions (\cite{SRW}) to results
from Monte-Carlo simulations (\cite{Ellis-etal}; \cite{Stecker-Tylka};
\cite{Jungman-Kam}; \cite{Bottino-etal}; \cite{Bottino-H}).  The
latter method has also been used together with $\bar p$ production
from neutralino annihilation in a minimal supergravity scheme
(\cite{Diehl-etal}).

In this work, we use the Lund Monte Carlo {\sc Pythia} 6.115
(\cite{pythia}) to simulate the energy spectrum of antiprotons from
neutralino annihilation. We have used large numerical tables for our
computations, but for convenience we also present useful
parametrisations of the $\bar p$ fluxes from various annihilation
channels. In the following Sections, we will describe the MSSM model
we use, describe the Monte
Carlo simulations, 
discuss the antiproton propagation model, 
discuss the background fluxes and the uncertainties
in both the background and the signal. Finally, we will show and
discuss our results.

\section{Definition of the Supersymmetric model}
\label{sec:MSSMdef}

We work in the Minimal Supersymmetric Standard Model (MSSM)\@.  In 
general, the MSSM has many free parameters, but with some reasonable 
assumptions we can reduce the number of parameters to the Higgsino 
mass parameter $\mu$, the gaugino mass parameter $M_{2}$, the ratio of 
the Higgs vacuum expectation values $\tan \beta$, the mass of the 
$CP$-odd Higgs boson $m_{A}$ (or $m_{H_3^0}$), the scalar mass 
parameter $m_{0}$ and the trilinear soft SUSY-breaking parameters 
$A_{b}$ and $A_{t}$ for the third generation.  In particular, we do 
not impose any restrictions from supergravity other than gaugino mass 
unification, which relates the other gaugino mass parameter $M_1$ to 
$M_2$.  (We remind that one of the most attractive features of the 
MSSM is that, unlike the non-supersymmetric Standard Model, it is 
compatible with gauge coupling unification given the current data on 
the running of low-energy gauge couplings.)  For a more detailed 
definition of the parameters and a full set of Feynman rules we refer 
to (\cite{coann}; \cite{jephd}).

The lightest stable supersymmetric particle is in most models the 
lightest neutralino (which we will henceforth just call `the 
neutralino', $\chi$), which is a superposition of the superpartners of 
the gauge and Higgs fields,
\begin{equation}
\chi\equiv  \tilde{\chi}^0_1 =
  N_{11} \tilde{B} + N_{12} \tilde{W}^3 +
  N_{13} \tilde{H}^0_1 + N_{14} \tilde{H}^0_2.
\end{equation}
It is convenient to define the gaugino fraction of the neutralino,
\begin{equation}
  Z_g = |N_{11}|^2 + |N_{12}|^2.
\end{equation}
For the masses of the neutralinos and charginos we use the one-loop
corrections from the literature
(\cite{neuloop}; \cite{neuloop2}; \cite{neuloop2a}; \cite{neuloop3})
and for the Higgs boson
masses we use the leading logarithmic two-loop radiative corrections,
calculated within the effective potential approach given by
\cite{carena}.

\begin{table}[!t]
  \begin{center}
  \begin{tabular}{rrrrrrrr} \hline
  Parameter & $\mu$ & $M_{2}$ &
  $\tan \beta$ & $m_{A}$ & $m_{0}$ & $A_{b}/m_{0}$ & $A_{t}/m_{0}$ \\
  Unit & GeV & GeV & 1 & GeV & GeV & 1 & 1 \\ \hline
  Min & -50000 & -50000 & 1.0  & 0     & 100   & -3 & -3 \\
  Max & 50000  & 50000  & 60.0 & 10000 & 30000 &  3 &  3 \\  \hline
  \end{tabular}
  \end{center}
\caption{The ranges of parameter values used in our scans of the MSSM
parameter space. Note that several special scans aimed at interesting
regions of the parameter space has been performed.  In total we have
generated approximately 116\,000 models which obey
all accelerator constraints. Of these, about 41000 have a relic
density in the range $0.025 < \Omega_\chi h^2 <1$\@.}
\label{tab:scans}
\end{table}

We make extensive scans of the model parameter space, some general and
some specialized to interesting regions.  In total we make 22
different scans of the parameter space.  The scans are done randomly
and are mostly distributed logarithmically in the mass parameters and
in $\tan \beta$\@.  For some scans the logarithmic scan in $\mu$ and
$M_{2}$ has been replaced by a logarithmic scan in the more physical
parameters $m_{\chi}$ and $Z_{g}/(1-Z_{g})$ where $m_{\chi}$ is the
neutralino mass. Combining all the scans, the overall limits of the
seven MSSM parameters we use are given in Table~\ref{tab:scans}.

We check each model to see if it is excluded by the most recent
accelerator constraints, of which the most important ones are the LEP
bounds (\cite{lepbounds}) on the lightest chargino mass,
\begin{equation}
  m_{\chi_{1}^{+}} > \left\{ \begin{array}{lcl}
  91 {\rm ~GeV} & \quad , \quad & | m_{\chi_{1}^{+}} -
  m_{\chi^{0}_{1}} |
  > 4 {\rm ~GeV} \\
  85 {\rm ~GeV} & \quad , \quad & {\rm otherwise}
  \end{array} \right.
\end{equation}
and on the lightest Higgs boson mass $m_{H_{2}^{0}}$ (which range from
72.2--88.0 GeV depending on $\sin (\beta-\alpha)$ with $\alpha$ being
a mixing angle in the Higgs sector) and the constraints from $b \to s \gamma$
(\cite{cleo}; \cite{cleo2}).  The new higher-precision measurement
from CLEO (Glenn \& al. 1998) gives a slightly smaller range for that
process than the one we have allowed; we have checked, however, that this 
causes no major changes in the properties related to $\bar p$ yield
for the allowed models. 

For each allowed model we compute
 the relic density of
neutralinos $\Omega_\chi h^2$, where $\Omega_\chi$ is the density in
units of the critical density and $h$ is the present Hubble constant
in units of $100$ km s$^{-1}$ Mpc$^{-1}$\@.  We use the formalism of
\cite{GondoloGelmini} for resonant annihilations, threshold
effects, and finite widths of unstable particles and we include all
two-body tree-level annihilation channels of neutralinos.  We also
include the so-called coannihilation processes according to
the results of Edsj\"o \& Gondolo (1997) in the relic density
calculation.

Present observations favor $h=0.6\pm 0.1$, and a total matter density
$\Omega_{M}=0.3\pm 0.1$, of which baryons may contribute 0.02 to 0.08
(see, e.g., \cite{cosmparams}).  Not to be overly restrictive, we accept
$\Omega_\chi h^2$ in the range from $0.025$ to $1$ as cosmologically
interesting.  The lower bound is
somewhat arbitrary as there may be several different components of
non-baryonic dark matter, but we demand that neutralinos are at least
as abundant as required to make up the dark halos of galaxies.  In
principle, neutralinos with $\Omega_\chi h^2<0.025$ would still be
relic particles, but only making up a small fraction of the dark
matter of the Universe.  We will consider models with $\Omega_\chi
h^2<0.025$ only when discussing the dependence of the signal on
$\Omega_\chi h^2$\@.

It may also be of interest to consider specifically models which naturally give
a value of $\Omega_\chi h^2$ which is close to the present `best fit' value. We
will therefore present some Figures where the range between 0.1 and 0.2 for
$\Omega_\chi h^2$ is shown by special symbols. Since in general a small
relic density implies a large annihilation cross section and vice versa, this
tends to cut out a large fraction of the models with otherwise observable
rates -- a fact not always highlighted in previous analyses.

\section{Antiproton production by neutralino annihilation}

\subsection{Introduction}

Neutralinos are Majorana fermions and
will annihilate with each other in the halo producing
leptons, quarks, gluons, gauge bosons and Higgs bosons. The quarks,
gauge bosons and Higgs bosons will decay and/or form jets that will
give rise to antiprotons (and antineutrons which decay shortly to
antiprotons).

At tree level the relevant final states for $\bar{p}$ production will
be $q \bar{q}$, $\ell \bar{\ell}$, $W^{+} W^{-}$, $Z^{0} Z^{0}$,
$W^{+} H^{-}$, $Z H_{1}^{0}$, $Z H_{2}^{0}$, $H_{1}^{0} H_{3}^{0}$ and
$H_{2}^{0} H_{3}^{0}$. We will include all the heavier quarks ($c$,
$b$ and $t$), gauge bosons and Higgs boson final states in our
analysis. In addition, we will include the $Z \gamma$ (\cite{ub}) and 
the 2 gluon (\cite{bua}; \cite{bu}) final states which occur at one 
loop-level. Note that for the antiproton-rich 2-gluon final state the 
improved and corrected formulas given in the second reference generally 
imply a lower branching ratio than those of the former reference
which has been used in several previous analyses.

The hadronization for all final states (including gluons) is simulated with
the well-known
particle physics Lund Monte Carlo program {\sc
Pythia} 6.115 (\cite{pythia}), which is used extensively at accelerators
in simulations of jet production at the full energy range which we need
to consider here. A word of caution should be raised, however, that
antiproton data is not very abundant, in particular not at the
lowest antiproton lab energies which tend to dominate our signal. 
Therefore an uncertainty in normalization, probably of the order of a 
factor 2, cannot be excluded at least in the low energy region.

\subsection{Simulations}
\label{sec:sim}

To get the energy distribution of antiprotons for each of the final
states listed in the previous subsection we generate the final states
$c \bar{c}$, $b \bar{b}$, $t \bar{t}$, $W^{+}W^{-}$, $Z^{0} Z^{0}$ and
$g g$ and let them decay/hadronize according to {\sc Pythia} 6.115.  We do not
need to include lighter quarks since the branching ratios to these are
negligible.  The annihilation channels containing Higgs bosons need
not be simulated separately since they decay to the other particles
for which we do simulate.  They are then let to decay in flight and
the spectrum from the decay products are boosted and averaged over the
decay angles.  During the simulations, antineutrons are let to decay
since we would otherwise underestimate the flux by a factor of two.

We have performed simulations for the neutralino masses $m_{\chi}$ =
10, 25, 50, 80.3, 91.2, 100, 150, 176, 200, 250, 350, 500, 750, 1000,
1500, 2000, 3000 and 5000 GeV and for intermediate masses an
interpolation is used. For each mass and annihilation
channel, $2.5\times 10^{5}$ events have been simulated.

\begin{table}[!t]
\begin{center}
\begin{small}
\begin{tabular}{cccccccccccccccc} \hline
 & \multicolumn{4}{c}{$p_1$} & & \multicolumn{2}{c}{$p_2$} & &
 \multicolumn{2}{c}{$p_3$} & & \multicolumn{4}{c}{$p_4$} \\
 \cline{2-5} \cline{7-8} \cline{10-11} \cline{13-16} 
 Channel & $a_{11}$ & $a_{12}$ & $a_{13}$ & $a_{14}$ & & 
 $a_{21}$ & $a_{22}$ & &
 $a_{31}$ & $a_{32}$ & &
 $a_{41}$ & $a_{42}$ & $a_{43}$ & $a_{44}$ \\ \hline
 $c \bar{c}$ &  1.70 & 1.40 & 0 & 0 & &
  3.12 & 0.04 & &
  $-2.22$ & 0 & &
  $-0.39$ & $-0.076$ & 0 & 0 \\
 $b \bar{b}$ & 1.75 & 1.40 & 0 & 0 & &
  1.54 & 0.11 & & 
  $-2.22$ & 0 & &
  $-0.31$ & $-0.052$ & 0 & 0 \\
 $t \bar{t}$ & 1.35 & 1.45 & 0 & 0 & &
  1.18 & 0.15 & & 
  $-2.22$ & 0 & & 
  $-0.21$ & 0 & 0 & 0 \\
  $W^+ W^-$ & 306.0 & 0.28 & $7.2\times10^{-4}$ & 2.25 & &
  2.32 & 0.05 & & 
  $-8.5$ & $-0.31$ & & 
  $-0.39$ & $-0.17$ & $-2.0\times10^{-2}$ & 0.23 \\
 $Z^0 Z^0$ & 480.0 & 0.26 & $9.6\times10^{-4}$ & 2.27 & &
  2.17 & 0.05 & &
  $-8.5$ & $-0.31$ & & 
  $-0.33$ & $-0.075$ & $-1.5\times10^{-4}$ & 0.71 \\
 $g g$ & 2.33 & 1.49 & 0 & 0 & &
  3.85 & 0.06 & & 
  $-2.17$ & 0 & & 
  $-0.312$ & $-0.053$ & 0 & 0 \\ \hline
\end{tabular}
\end{small}
\end{center}
\caption{The parameters fitted to the antiproton distributions for 
neutralino masses 50--5000 GeV\@.  The parameters $a_{23}$, $a_{24}$, 
$a_{33}$ and $a_{34}$ are always zero and not given in the table.
The parameterized distributions are given by 
Eqs.~(\protect\ref{eq:pbparam})--(\protect\ref{eq:ppoly}).}
\label{tab:param}
\end{table}

\begin{figure}
\centerline{\epsfig{file=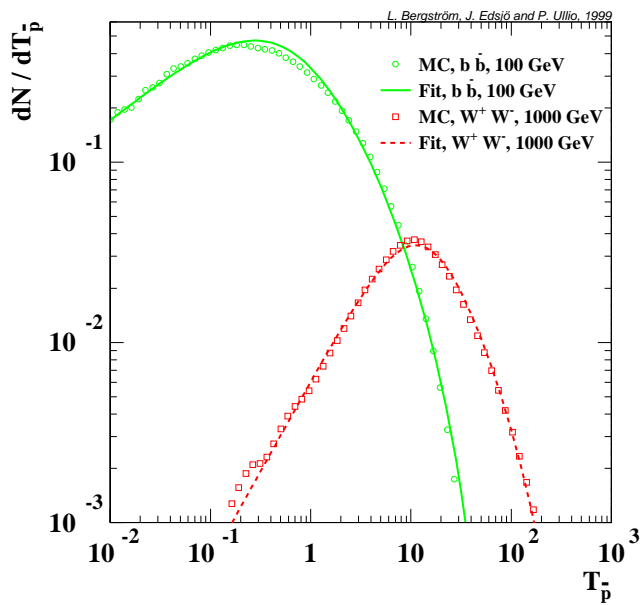,width=0.49\textwidth}}
\caption{Examples of the fits to the antiproton distributions from
neutralino annihilation, $dN/dT=(1/m_{\chi})dN/dx$ with $dN/dx$ given 
by Eqs.~(\protect\ref{eq:pbparam})--\protect\ref{eq:ppoly}) and 
Table~\protect\ref{tab:param}. }
\label{fig:pbanp}
\end{figure}

For easier use, we have also parameterized the antiproton
distributions for the `basic' annihilation channels given above. It is
for this purpose more convenient to define
$x=T_{\bar{p}}/m_{\chi}$, with $T_{\bar{p}}$ being the
kinetic energy of the antiproton, as the independent variable. A
suitable parameterization is then given by
\begin{equation}
  \frac{dN}{dx} = \left( p_1 x^{p_3} + p_2 \left | \log_{10} x
  \right|^{p_4} \right)^{-1}
  \label{eq:pbparam}
\end{equation}
where the parameters $p_i$ depend on both annihilation channel and the
neutralino mass. The latter dependence is parameterized as
\begin{equation}
  p_i(m_\chi) = \left(a_{i1} m_\chi^{a_{i2}} + a_{i3} m_\chi^{a_{i4}}
  \right)^{-1} 
  \label{eq:ppoly}
\end{equation}
The values of the
$a_{ij}$ for the different annihilation channels are given in
Table~\ref{tab:param}. These parameterizations are valid for
neutralino masses in the range 50--5000 GeV. The error in the
most relevant regions, i.e.\ the low-energy tail to most of the
high-energy slope is usually less than 20\%. At worst (for $t\bar{t}$),
it can be up to 50\% in isolated regions.
This should be compared with the
uncertainties in {\sc Pythia} which probably can be up to a factor of 2.
In Fig.~\ref{fig:pbanp} we give as an example the distributions for a
100 GeV neutralino annnihilating into $b\bar{b}$ and a 1000 GeV
neutralino annihilating into $W^+W^-$.

Note that the above parameterizations are only given for the
reader's convenience - in our calculations we use the results of the
simulations directly. For convenience we also show in
Table~\ref{tab:examples}  the individual branching ratios of the main
modes for a set of models with high antiproton yield but different mass,
and gaugino content.

\begin{table}[!t]
  \begin{small}
  \begin{center}
  \begin{tabular}{lccccccc} \hline
%
%
Example No.\ & 1 & 2 & 3 & 4 & 5 & 6 & 7  \\ \hline
%
%
$\mu$        & $926.6$ & $-191.2$ & $708.7$   & $-127.6$ &
  $-543.1$   & $-891.2$ & $379.6$ \\ 
$M_2$        & $92.4$  & $-149.8$ & $-2156.1$ & $242.3$  &
  $413.3$    & $111.5$ & $-162.4$ \\ 
$\tan \beta$ & $19.1$  & $45.1$   & $9.67$    & $5.09$   &
  $4.17$     & $38.6$ & $28.8$ \\ 
$m_A$        & $79.8$  & $543.3$  & $1405.0$  & $1838.5$ &
  $409.2$    & $77.9$ & $89.5$ \\ 
$m_0$        & $922.5$ & $306.6$  & $2817.0$  & $4243.2$ &
  $1054.6$   & $1110.6$ & $1109.1$ \\ 
$A_b/m_0$    & $0.95$  & $0.17$   & $-2.05$   & $1.22$   &
  $-1.28$    & $0.17$ & $1.84$\\ 
$A_t/m_0$    & $-2.11$ & $0.96$   & $0.46$    & $-0.38$  &
  $-1.72$    & $2.06$ & $-2.20$\\ \hline
%
%
$m_{\chi^0}$ & $46.2$    & $69.4$   & $704.9$   & $98.9$    &
  $208.2$    & $56.1$ & $80.9$ \\
$Z_g$        & $0.99761$ & $0.88843$& $0.00642$ & $0.37943$ &
  $0.99293$  & $0.99756$ & $0.98440$ \\
$m_{\chi^+}$ & $91.0$    & $117.9$  & $708.3$   & $123.8$   &
  $405.8$    & $110.9$ & $155.7$ \\
$m_{H_1^0}$  & $126.6$   & $543.3$  & $1405.2$  & $1838.9$  &
  $412.0$    & $126.8$ & $127.4$ \\
$m_{H_2^0}$  & $76.7$    & $107.5$  & $117.8$   & $115.5$   &
  $115.5$    & $76.2$ & $89.0$ \\
$m_{H^\pm}$  & $111.5$   & $549.7$  & $1407.1$  & $1840.1$  &
  $416.4$    & $115.5$ & $120.0$ \\ \hline
%
%
$(\sigma_{\rm ann} v)$ & 0.127 & 0.0874 & 0.420 & 0.0719 & 0.116 & 
0.0824 & 0.0785 \\
$Br(c \bar{c})$ & $<10^{-5}$ & $0.00001$ & 
  $<10^{-5}$ & $0.00050$ &
  $0.00007$ & $<10^{-5}$ & $<10^{-5}$ \\
$Br(b \bar{b})$ & $0.96$ & $0.98$ &
  $0.81$ & $0.0059$ &
  $0.32$ & $0.96$ & $0.96$ \\
$Br(t \bar{t})$ & $-$  & $-$  & 
  $0.11$ & $-$  &
  $0.65$ & $-$ & $-$ \\
$Br(W^+ W^-)$   & $-$  & $-$  & 
  $0.024$ & $0.87$ &
  $<10^{-5}$ &  $-$ & $<10^{-5}$ \\
$Br(Z^0 Z^0)$   & $-$  & $-$  &
  $0.020$ & $0.12$ &
  $<10^{-5}$ & $-$ & $-$ \\
$Br(gg)$        & $0.00060$ & $0.0028$ &
  $0.00017$ & $0.0016$ &
  $0.0028$ & $0.00054$ & $0.00043$ \\
$Br(\gamma \gamma)$ & $3.7\times10^{-7}$  & $3.9\times10^{-6}$ & 
  $6.5\times10^{-5}$ & $1.8\times10^{-4}$ & $2.3\times10^{-6}$ & 
  $9.6\times10^{-7}$ & $1.2\times10^{-6}$ \\
$Br(Z \gamma)$      & $5.8\times10^{-13}$ & $5.8\times10^{-6}$ &
  $1.7\times10^{-4}$ & $6.5\times10^{-4}$ & $2.3\times10^{-5}$ & 
  $6.3\times10^{-10}$ & $2.1\times10^{-6}$ \\ \hline
%
%
$\Omega_\chi h^2$ & 0.025 & 0.028 & 0.032 & 0.025 & 
  0.050 & 0.034 & 0.034 \\
$\Phi_{\bar{p}}$ & $4.6\times10^{-2}$ & $1.6\times10^{-2}$ &
  $8.5\times10^{-4}$  & $8.8\times10^{-3}$ &
  $4.0\times10^{-3}$ & $2.1\times10^{-2}$ & $1.1\times10^{-2}$ \\
$\Phi_{\mu,\rm~Earth}$ & $3.8\times10^{-1}$ & $1.0\times10^{0}$ &
  $2.4\times10^{-5}$ & $1.1\times10^{-2}$ & $2.5\times10^{-7}$ &
  $7.5\times10^{1}$ & $2.3\times10^{1}$ \\
$\Phi_{\mu,\rm~Sun}$ & $7.0\times10^{-2}$ & $2.6\times10^{0}$ &
  $4.1\times10^{-1}$ & $2.7\times10^{3}$ &
  $8.7\times10^{-1}$ & $1.3\times10^{0}$ & $1.8\times10^{1}$ \\
$\sigma_{\rm SI}$    & $2.9\times10^{-7}$ & $4.0\times10^{-7}$ &
  $9.6\times10^{-10}$  & $1.2\times10^{-8}$ &
  $1.2\times10^{-10}$ & $1.2\times10^{-6}$ & $2.3\times10^{-6}$ \\
$\Phi_{e^+}$ & $5.0\times10^{-8}$ & $2.9\times10^{-8}$ &
  $3.2\times10^{-8}$ & $2.6\times10^{-8}$ &
  $2.0\times10^{-8}$ & $3.1\times10^{-8}$ & $2.6\times10^{-8}$ \\
$\Phi_{{\rm cont.}~\gamma}$ & $8.1\times10^{-8}$ & $3.5\times10^{-8}$ &
  $5.9\times10^{-9}$ & $1.4\times10^{-8}$ &
  $1.1\times10^{-8}$ & $4.2\times10^{-8}$ & $2.6\times10^{-8}$ \\ 
  \hline
\end{tabular}
\end{center}
\end{small}
\caption{Example of models giving high antiproton rates.  All masses
  are given in GeV; the annihilation rate is given in $10^{-24}$
  cm$^{3}$ s$^{-1}$; the solar modulated $\bar{p}$ flux at 0.35 GeV is
  in m$^{-2}$ s$^{-1}$ sr$^{-1}$ sr$^{-1}$; the neutrino flux from the
  Earth and the Sun is with a threshold of 25 GeV and in units of
  km$^{-2}$ yr$^{-1}$; the spin-independent cross section is in pb;
  the solar modulated positron flux is in one of the HEAT
  bins (8.9-14.8 GeV) and in units of cm$^{-2}$ s$^{-1}$ sr$^{-1}$
  GeV$^{-1}$; the continuum gamma flux is for high Galactic latitudes
  and is given in cm$^{-2}$ s$^{-1}$ sr$^{-1}$, integrated above 1
  GeV. The rates and fluxes are calculated as given in
  (\protect\cite{bg}; \protect\cite{clumpy}; \protect\cite{eplus}; 
  \protect\cite{bub}; \protect\cite{beg}). }
\label{tab:examples}
\end{table}


\subsection{The Antiproton Source Function}

The source function $Q_{\bar{p}}^{\chi}$ gives the number of antiprotons 
per unit time, energy and volume element produced in annihilation
of neutralinos locally in space. It is given by
\begin{equation}
  Q_{\bar{p}}^{\chi}(T,\vec{x}\,)=
  (\sigma_{\rm ann}v)
  \left(\frac{\rho_{\chi}(\vec{x}\,)}{m_{\chi}}\right)^{2}
  \sum_{f}^{}\frac{dN^{f}}{dT}B^{f}
  \label{sourcefkn}
\end{equation}
where $T$ is the $\bar{p}$ kinetic energy. For a given annihilation
channel $f$, $B^{f}$ and $dN^{f} / dT$ are, respectively, the branching ratio
and the fragmentation function, and $(\sigma_{\rm ann}v)$ is the
annihilation rate at $v=0$ (which is very good approximation since the
velocity of the neutralinos in the halo is so low).
As dark matter neutralinos annihilate in pairs, the source function is
proportional to the square of the neutralino number density 
$n_{\chi} = \rho_{\chi} / m_{\chi}$. 
Assuming that most of the dark matter in the Galaxy is made up of neutralinos
and that these are smoothly distributed in the halo, one can directly relate
the neutralino number density to the dark matter density profile in the 
galactic halo $\rho$. Given a generic parametrization of $\rho$, we fix:  
\begin{equation}
  \rho_{\tilde{\chi}}(\vec{x}\,) \equiv \rho(\vec{x}) =
  \rho_0\; \left(\frac{r_0}{|\vec{x}|}\right)^{\gamma}
  \,\left[\frac{1+(r_0/a)^\alpha}{1+(|\vec{x}|/a)^\alpha}\right]
  ^{(\beta-\gamma)/\alpha}
\label{eq:haloprof}
\end{equation}
where $\rho_0$ is the value of the local halo density, $r_0$ is the 
galactocentric distance of the Sun and $a$ is some length scale;
we assume $\rho_0 = 0.3\ \rm{GeV} \rm{cm}^{-3}$ and $r_0 = 8.5\ \rm{kpc}$.
In the actual computation we will mainly restrict ourselves
to the case in which the dark
matter density profile is described by a modified isothermal distribution,
$(\alpha,\beta,\gamma)=(2,2,0)$, mentioning what changes are expected in case 
more cuspy profiles, which are favoured by results in $N$-body simulations 
of hierarchical clustering, are considered. We will in particular consider 
the example of the Navarro et al.\ profile (\cite{navarro}), 
$(\alpha,\beta,\gamma)=(1,2,1)$.
Although we are here focusing on the case of a smooth distribution of dark 
matter particles in the halo, an extension to a clumpy distribution is 
potentially interesting as well (\cite{clumpy}; \cite{pieroclumpy}).

Given the $\bar{p}$ distributions calculated in the previous section, we can 
now get the source function for any given annihilation channel.

\section{Propagation model}
\label{sec:prop}

In the absence of a well established theory to describe the
interactions of charged particles with the magnetic field of the
Galaxy and the interstellar medium, the propagation of cosmic rays has
generally been treated by postulating a semiempirical model and
fitting the necessary set of unknown parameters to available data.  A
common approach is to use a diffusion approximation defined by a
transport equation and an appropriate choice of boundary
conditions~(see e.g. \cite{Berezinskii}; \cite{Gaisserbook} and
references therein).

In most diffusion models the form of the terms present in the
diffusion equation is a compromise between physical insight and the
possibility of an analytical solution.  Only
recently more realistic models have been studied by applying numerical
solutions~(\cite{StrongMoskalenko}) or Monte Carlo
simulations~(\cite{PorterProtheroe}).

Our analysis is focused on comparing the characteristics of
cosmic-ray antiproton signals of different origin: secondary
antiprotons produced in cosmic-ray interactions and, eventually, a
primary flux from neutralino annihilations. We want in particular to
examine the dependence of the relative strength and spectral
signatures on the diffusion model and the distribution of particle
dark matter in the galactic halo.  As both of them are not well
constrained, we believe that an analytic solution of a reasonable
physical model will be sufficient to provide most of the information
needed on the behaviour of the two types of signals.

We choose to describe the propagation of cosmic rays in the Galaxy by a
 transport equation of the diffusion type as written by Ginzburg and
Syrovatskii (1964) (see also
\cite{Berezinskii}; \cite{Gaisserbook}).
In the case of a stationary solution, the number density $N$ of a stable cosmic ray
species whose distribution of sources is defined by the function of
energy and space $Q(E,\vec{x})$, is given by:
\begin{equation}
\frac{\partial{N(E,\vec{x})}}{\partial{t}} = 0 = \nabla \cdot
\left(D(R,\vec{x})\,\nabla N(E,\vec{x})\right)
- \nabla \cdot \left( \vec{u}(\vec{x})\,N(E,\vec{x}) \right)
- p(E,\vec{x})\,N(E,\vec{x}) + Q(E,\vec{x}) \;\;.
\label{eq:diff}
\end{equation}
Here and below we try to keep the notation as general as possible.
Although our goal is to compute $N$ for antiprotons, in order to determine
the source function for the secondary flux we will have to obtain the
spatial density distribution for protons as well. On the right hand side
of Eq.~(\ref{eq:diff}) the first term implements the diffusion approximation
for a given diffusion coefficient $D$, generally assumed to be a function of
rigidity $R$, while the second term describes a large-scale convective
motion of velocity $\vec{u}$. The third term is added to take into account
losses of cosmic rays due to to collisions with the interstellar matter.
It is a very good approximation to include in this term only the interactions
with interstellar hydrogen (on the other hand, we will point out below 
that heavier elements, in particular helium, cannot be neglected when
computing the source function for secondary antiprotons); in this case
$p$ is given by:
\begin{equation}
p(E,\vec{x}) = n^H(\vec{x}) \, v(E) \, \sigma^{\rm in}_{cr\,p}(E)
\end{equation}
where $n^H$ is the hydrogen number density in the Galaxy, $v$ is the
velocity of the cosmic ray particle considered `$cr$', while
$\sigma^{\rm in}_{cr\,p}$ is the inelastic cross section for $cr$-proton
collisions. In Eq.~(\ref{eq:diff}) we have neglected continuous energy losses;
this will be included in an implicit form when considering secondary
antiprotons.
We will briefly mention in the conclusions the possibility that antiprotons 
have a finite lifetime $\tau$. To take this effect into account the term
$- (1/\tau)\, N(E,\vec{x})$ should be added on the right hand side of 
Eq.~(\ref{eq:diff}), and this corresponds to shifting $p$ to $p + 1/\tau$
in all the equations below.

We now have to specify the parameters introduced and the boundary
conditions. We mainly follow the approach of \cite{GKP}, given also in
\cite{Berezinskii} and analogous to that of \cite{WLG},
\cite{Chardonnet}, and \cite{bottinolast}. The main feature is that
the propagation region is assumed to have a cylindrical symmetry: the
Galaxy is split into two parts, a disk of radius $R_h$ and height
$2\cdot h_g$, where most of the interstellar gas is confined, and a
halo of height $2\cdot h_h$ and the same radius. We assume that the
diffusion coefficient is isotropic with possibly two different values
in the disk and in the halo, reflecting the fact that in the disk
there may be a larger random component of the magnetic fields. We then
have a spatial dependence:
\begin{equation}
D(\vec{x}) = D(z) = D_g \, \theta(h_g - |z|) + D_h \, \theta(|z| - h_g)\;\;.
\end{equation}
Regarding the rigidity dependence, fits to cosmic ray data in models which
do not include reacceleration effects indicate that $D$ scales as
$R^{0.6}$~(\cite{WLG}; \cite{StrongMoskalenko}) with a cutoff below some rigidity $R_0$.
We consider the same functional form as in
\cite{Chardonnet} and \cite{bottinolast}:
\begin{equation}
D_l(R) = D_l^0 \left(1+\frac{R}{R_0}\right)^{0.6}
\label{eq:diffco}
\end{equation}
where $l=g,h$.
We will briefly discuss below what changes are expected in case
reacceleration is included, without making numerical predictions.
The convective term has been introduced in Eq.~(\ref{eq:diff}) to describe the
effect of particle motion against the wind of cosmic rays leaving the disk.
We will therefore not  consider any convection in the radial direction,
assuming instead a galactic wind of velocity
\begin{equation}
\vec{u}(\vec{x}) = \left(0, 0, u(z)\right)
\end{equation}
where
\begin{equation}
u(z)= {\rm sign}(z) \, u_h \, \theta(|z| - h_g)\;\;.
\end{equation}
An analytic solution is possible also in the case of a linearly increasing
wind~(\cite{pieroclumpy}). The last parameter we have to specify is the
distribution of gas in the Galaxy: for convenience we assume that this has
the very simple z dependence
\begin{equation}
n^H(\vec{x}) = n^H(z)
= n_g^H \, \theta(h_g - |z|) + n_h^H \, \theta(|z| - h_g)
\end{equation}
where $n_h \ll n_g$ (in practice we will take $n_h=0$) and an average
in the radial direction is performed. Finally, as we eventually
want to treat the case of clumpy neutralino dark matter (\cite{pieroclumpy})
we will not assume any symmetry in the source function $Q(\vec{x})$
(note that to apply the results of this Section to sources with a cylindrical 
symmetry, as for instance Eq.~(\ref{sourcefkn}) with the assumption in 
Eq.~(\ref{eq:haloprof}), it is sufficient to set everywhere the index 
$k$ equal to 0).

As boundary condition, it is usually assumed that cosmic rays can escape
freely at the border of the propagation region, i.e.\
\begin{equation}
N(R_h,z) = N(r,h_h) = N(r,-h_h) = 0
\label{eq:bound}
\end{equation}
as the density of cosmic rays is assumed to be negligibly small in the
intergalactic space. To check whether this hypothesis holds even in the
case of a source from dark matter annihilations, we will 
compare the flux of outgoing antiprotons with the
one entering the diffusion region due to sources in `free space'.
For references about other possible choices of boundary conditions
see~\cite{Berezinskii}.

The cylindrical symmetry and the free escape at the boundaries makes it possible
to solve the transport equation expanding the number density distribution $N$
in a Fourier-Bessel series:
\begin{equation}
N(r,z,\theta) = \sum_{k=0}^{\infty} \,\sum_{s=1}^{\infty} \;
J_k \left(\nu_s^k \frac{r}{R_h}\right) \cdot
\left[ M_s^k(z) \cos(k\theta) + \tilde{M}_s^k(z) \sin(k\theta)\right] \\
\end{equation}
which automatically satisfies the boundary condition at $r=R_h$,
$\nu_s^k$ being the $s$-th zero of $J_k$ (the Bessel function of the
first kind  and of order $k$).
In the same way the source function can be expanded as:
\begin{equation}
Q(r,z,\theta) = \sum_{k=0}^{\infty} \,\sum_{s=1}^{\infty} \;
J_k \left(\nu_s^k \frac{r}{R_h}\right) \cdot
\left[ Q_s^k(z) \cos(k\theta) + \tilde{Q}_s^k(z) \sin(k\theta)\right]
\end{equation}
where
\begin{equation}
Q_s^k(z) = \frac{2}{{R_h}^2\,{J_{k+1}}^2(\nu_s^k)}
\int\limits_0^{R_h} dr^{\prime} \;r^{\prime}
J_k \left(\nu_s^k \frac{r^{\prime}}{R_h}\right)
\frac{1}{\alpha_k\,\pi}
\int\limits_{-\pi}^{\;\pi} d\theta^{\prime} \; \cos(k\theta^{\prime})
\,Q(r^{\prime},z,\theta^{\prime})\;\;.
\end{equation}
In the equation above $\alpha_0=2$, while $\alpha_k=1$ for $k\geq 1$; it is
not necessary to specify the coefficients of the terms in $\sin(k\theta)$
as we fix the coordinate system such that $\theta=0$ at our location
and we are only interested in computing fluxes for this value of $\theta$.
Inserting the two expansions in Eq.~(\ref{eq:diff}), we can derive the
equation relevant for the propagation in the z direction:
\begin{equation}
\frac{\partial}{\partial{z}} D(z) \frac{\partial}{\partial{z}} M_s^k(z)
- D(z) \left(\frac{\nu_s^k}{R_h}\right)^2 M_s^k(z)
- \frac{\partial}{\partial{z}} \left(u(z)\,M_s^k(z) \right)
- p(z) M_s^k(z) + Q_s^k(z) =0 \;\;.
\end{equation}
The solution of this equation is straightforward: it can be easily derived
writing a solution separately for $h_g<z<h_h$, $-h_g<z<h_g$ and $-h_h<z<-h_g$,
and then imposing the boundary conditions at $z=\pm h_h$,
i.e.\ Eq.~(\ref{eq:bound}), and the continuity of the number density and of
the flux, that is of
$M_s^k(z)$ and $[-D(z) \partial/\partial{z}\,M_s^k(z)+u(z)\,M_s^k(z)]$,
at $z=\pm h_g$. For $-h_g \leq z \leq h_g$ the solution is given by:
\begin{equation}
M_s^k(z) = M_s^k(0) \cosh(\lambda_g^{ks} z) - \frac{1}{D_g\, \lambda_g^{ks}}
\int\limits_{0}^{z} \;dz^{\prime}
\sinh\left(\lambda_g^{ks} (z-z^{\prime})\right) Q_s^k(z^{\prime})
\label{eq:zdep}
\end{equation}
where
\begin{eqnarray}
M_s^k(0) \ = \ \frac{1}{\cosh(\lambda_g^{ks} h_g)}
\left\{\frac{I_H}{\sinh\left(\lambda_h^{ks} (h_h-h_g)\right)}+
\frac{D_h\,I_{GS}}{D_g\, \lambda_g^{ks}} \,\left[\gamma_h +
\lambda_h^{ks} \coth\left(\lambda_h^{ks} (h_h-h_g)\right)\right]
+I_{GC}\right\} \nonumber \\
\ \ \times \left[D_g\,\lambda_g^{ks}\,\tanh\left(\lambda_g^{ks} h_g\right)
+ D_h\,\gamma_h + D_h\,\lambda_h^{ks}\,
\coth\left(\lambda_h^{ks} (h_h-h_g)\right) \right]^{-1}
\label{eq:halodiff}
\end{eqnarray}
and we have introduced the following set of definitions:
\begin{eqnarray}
\lambda_g^{ks} = \sqrt{\left({\nu_s^k\over R_h}\right)^2 +
\frac{n_g^H v \sigma^{\rm in}_{cr\,p}}{D_g}} \;,\;\;\;\;\
\lambda_h^{ks} = \sqrt{\left({\nu_s^k\over R_h}\right)^2 +
\frac{n_h^H v \sigma^{\rm in}_{cr\,p}}{D_h} + {\gamma_h}^2}\;,\;\;\;\; \
\gamma_h = \frac{u_h}{2\,D_h}
\label{eq:halodiff2}
\end{eqnarray}
and
\begin{eqnarray}
I_H & = & \int\limits_{h_g}^{h_h} \;dz^{\prime} \,
\sinh\left(\lambda_h^{ks} (h_h-z^{\prime})\right) \,
\exp\left(\gamma_h (h_g-z^{\prime})\right)
\cdot \frac{Q_s^k(z^{\prime}) + Q_s^k(-z^{\prime})}{2} \nonumber \\
I_{GS} & = & \int\limits_{0}^{h_g} \;dz^{\prime} \,
\sinh\left(\lambda_g^{ks} (h_g-z^{\prime})\right)
\cdot \frac{Q_s^k(z^{\prime}) + Q_s^k(-z^{\prime})}{2} \nonumber \\
I_{GC} & = & \int\limits_{0}^{h_g} \;dz^{\prime} \,
\cosh\left(\lambda_g^{ks} (h_g-z^{\prime})\right)
\cdot \frac{Q_s^k(z^{\prime}) + Q_s^k(-z^{\prime})}{2} \;.
\label{eq:int}
\end{eqnarray}
For $|z| > h_g$ the solution is analogous, but less compact and will
not be reproduced here. It is more useful to give explicitly
$M_s^k(0)$ in case of a source which is constant in the disk and negligible
in the halo; in this case
\begin{eqnarray}
M_s^k(0) & = & \frac{1}{D_g\,{\lambda_g^{ks}}^2}
\left\{1 -
\frac{D_h\,\left[\gamma_h + \lambda_h^{ks}
\coth\left(\lambda_h^{ks} (h_h-h_g)\right)\right]}
{D_g\,\lambda_g^{ks}\,\sinh\left(\lambda_g^{ks} h_g\right)
+ D_h\,\cosh\left(\lambda_g^{ks} h_g\right)
\left[\gamma_h + \lambda_h^{ks}
\coth\left(\lambda_h^{ks} (h_h-h_g)\right)\right]} \right\}
\; Q_s^k(0) \nonumber \\
&\equiv & M^*(s,k)\; Q_s^k(0) \;.
\label{eq:condiff}
\end{eqnarray}
It is easy to check that in the limit $u_h \rightarrow 0$,
Eq.~(\ref{eq:condiff})
correctly reduces to the result quoted in~\cite{Berezinskii}, Chapter~3,
Section~3. Also Eq.~(\ref{eq:halodiff}) corresponds to the number density found
in \cite{Chardonnet} and \cite{bottinolast} in the limits $h_g \rightarrow 0$,
$D_h \rightarrow D_g$, $u_h \rightarrow 0$ and for a source function symmetric
with respect to the z axis.

Before applying these results to compute cosmic ray antiproton fluxes,
let us pause for a moment. Most commonly, data on cosmic rays have
been treated within the framework of the leaky box approximation.
This is, to a certain extent, a simplified version of the diffusion
model, where it is assumed that diffusion takes place rapidly. As was
noticed for instance in \cite{Berezinskii}, the path length
distribution function of particles for a diffusion model with sources
in the disk is very close to the exponential form characteristic of
the leaky box treatment. For the purpose of computing secondary
antiprotons, which are mainly generated in the gaseous disk, we expect
it to be essentially equivalent to write a diffusion equation and fit
its parameters to existing data on cosmic ray nuclei, or to derive
from the same data in the simple leaky box scenario the grammage as a
function of rigidity and use it to compute the secondary antiproton
flux.

In this sense we do expect to find an antiproton secondary spectrum
which is analogous to the results of several papers in which this has
been calculated in the leaky box approximation (except that in some of
these papers not all the relevant effects have been included).  On the
other hand, in the case of neutralino sources which are more
homogeneously distributed extending through the full halo, it is
unlikely that the effective average matter density particles have gone
through can be of the same form as for sources located only in the
disk. We will try to analyze in some detail this dependence on the
geometry of the source, and we believe that this will give a real
improvement with respect to the leaky box approximation.


\section{Solar Modulation}

A further complication when comparing predictions of a theoretical
model with data on cosmic rays taken at Earth is given by the solar
modulation effect. During their propagation from the interstellar
medium through the solar system, charged particles are affected by the 
solar wind and tend to lose energy. The net result of the modulation
is a shift in energy between the interstellar spectrum and the 
spectrum at the Earth and a substantial depletion of particles with
non-relativistic energies.

The simplest way to describe the phenomenon is the analytical force-field 
approximation by Gleeson \& Axford (1967; 1968) for a spherically 
symmetric model. The prescription of 
this effective treatment is that, given an interstellar flux at the 
heliospheric boundary, $d\Phi_{\rm b}/dT_{\rm b}$, the flux at the Earth 
is related to this by
\begin{equation}
  \frac{d\Phi_{\oplus}}{dT_\oplus}(T_\oplus) = \frac{p_\oplus^2}{p_{\rm
  b}^2} \frac{d\Phi_{\rm b}}{dT_{\rm b}} (T_{\rm b})
\end{equation}
where the energy at the heliospheric boundary is given by
\begin{equation}
  E_{\rm b} = E_\oplus + |Ze|\phi_F
\end{equation}
and $p_{\otimes}$ and $p_{\rm b}$ are the momenta at the Earth and 
the heliospheric boundary respectively.
Here $e$ is the absolute value of the electron charge and $Z$ the
particle charge in units of $e$ (e.g. $Z=-1$ for antiprotons).

An alternative approach is to solve numerically the propagation equation 
of the spherically symmetric model~(\cite{fisk}): the solar modulation 
parameter one has to introduce with this method roughly corresponds to 
$\phi_F$ as given above. When computing solar modulated antiproton fluxes,
the two treatments seem not to be completely equivalent in the low energy 
regime (the reader may check for instance Fig.~4 in~(\cite{Labrador}) 
against Fig.~8 in~(\cite{bottinolast})); keeping this in mind, we will
anyway
implement the force field approximation, avoiding the problem of 
having to solve a partial differential equation for each of our 
supersymmetric models.

We just mention here that the picture can be much more complicated:
non-spherical propagation models which take into account the polarity
of the solar magnetic field have been studied as well (e.g.~\cite{WP}).
In this case the solar modulation effects on particles and antiparticles 
can be quite different. If one would translate this into the force field
effective treatment, one should use different values for the modulation 
parameter for protons and antiprotons (in contrast with the standard
procedure of assigning to antiprotons the value found from proton flux 
measurements). The relation between these two would be very model
dependent.

To compare with the two sets of {\sc Bess} measurements, which are both
near solar minimum, we choose $\phi_F=500$~MV, in reasonable
agreement with what the {\sc Bess} collaboration uses in their analysis.
Otherwise, we will focus on the interstellar fluxes which are not
affected by these uncertainties.

\section{Background estimates}

\subsection{General considerations}
Secondary antiprotons are produced in cosmic ray collisions with the
interstellar gas. Looking at the composition of incident and target
particles it is easy to guess that the main contribution to the
$\bar{p}$ flux is given by cosmic ray protons colliding with interstellar
hydrogen atoms. Because of baryon number conservation, the minimal
$p + p \to \bar{p} + X$ reaction has three protons in the final state;
in the rest frame of the target hydrogen atom, the energy threshold for an
incident proton to produce an antiproton is therefore $E_p = 7 m_p$.
Due to this feature, the energy distribution of the produced $\bar{p}$
shows a sharp peak at a few GeV and a steep fall-off at lower energies.
It is reasonable to expect that in this low energy region
reactions involving heavier nuclei, both as targets and projectiles,
may play some role: they imply in fact different kinematics and the
spectrum of the produced $\bar{p}$ need not fall as fast as for low
energy $pp$ collisions.
We have verified that the interaction of primary protons with interstellar
helium is indeed a relevant process, while all others can be safely
neglected as their contributions add up to below a few per cent of the
total
at any energy (this is essentially the same conclusion
as that of \cite{smr}, although our approach is slightly different, as we
will point out below).
We assume therefore that the source function for secondary antiprotons has
the following form (the factor of 2 accounts for antiprotons produced by
antineutron decays):
\begin{equation}
Q_{\bar{p}}(\vec{x},E) = 2\cdot 4\pi\; \int\limits_{E_{\rm thresh}}^{\infty}
dE^{\prime}
\; \left[ \frac{d\sigma_{pH\rightarrow \bar{p}}}{dE}(E,E^{\prime})
\,n^H(\vec{x}) + \frac{d\sigma_{pHe\rightarrow \bar{p}}}{dE}(E,E^{\prime})
\,n^{He}(\vec{x}) \right]\, \Phi_p(\vec{x},E^{\prime})\;.
\label{eq:source}
\end{equation}
In this formula $\Phi_p(\vec{x},E^{\prime})$ is the primary proton cosmic ray
flux at the position $\vec{x}$ in the Galaxy and for the energy
$E^{\prime}$, $n^{He}$ is the helium number density which we assume to be
$7 \%$ of $n^H$ (\cite{GarciaMunoz}) and
have the same spatial dependence, while $d\sigma /dE\,(E,E^{\prime})$
stands for the differential cross section for the production of an
antiproton with energy $E$ for an incident proton of energy $E^{\prime}$,
in the two processes considered.  For $pH$ collisions we implement the
standard parametrization for the differential cross section introduced
in \cite{tanng}; as already mentioned the energy threshold
$E_{\rm thresh}$ for this process is $E^{\prime} = 7 m_p$.

\subsection{Antiproton production in collisions with nuclei}

A much smaller set of data is available in the case of antiproton
production in proton collisions with heavier elements. In particular,
it is difficult to estimate the effects of production below the
nominal energy threshold for $p+p\to \bar p + X$, which is known to
occur in hadron-nucleus collisions. Recently, several experiments have
shown substantial sub-threshold $\bar p$ production for deuterium,
helium, carbon and copper targets (\cite{chiba}; \cite{schroter}).
Possible collective effects allow the abundant
low-energy cosmic rays to produce antiprotons below threshold.
On their way out of the nucleus, the produced antiprotons
may  also 
suffer inelastic losses which slow them down, creating a potentially
important component in the low-energy cosmic ray-induced $\bar p$
spectrum. (Note that helium and heavier nuclei in the cosmic rays also
give different kinematics for produced antiprotons. However, this gives
an extra contribution at higher $\bar p$ energies, and is therefore not
important for our study.)
These sub-threshold effects have been modeled by \cite{sibirtsev}, where
the limited data set available can be described by a transport equation
solved by a Monte Carlo technique. 

Here we follow a much simplified approach, which describes the C and
Cu data displayed in \cite{sibirtsev} reasonably well and which we apply
to $p$ + He $\to \bar p + X$. We find that 
the collective effects can be mimicked
by a shift in the incident proton energy:
\begin{equation}
E_{\rm in}\to E_{\rm eff}=E_{\rm in}+0.6(E_{\rm thresh}-E_{\rm in})
  \theta(E_{\rm thresh}-E_{\rm in})\
+1.1\ {\rm GeV},
\end{equation}
where $E_{\rm thresh}$ is the nominal threshold ($7m_p$) for $\bar p$
production in $pp$ collisions. The energy loss due to inelastic rescattering
can be approximated by decreasing the energy of the outgoing antiproton
(using $pp$ kinematics) by 1 GeV. The yield of antiprotons per
collision is taken to scale with the total $pA$ cross
section, parametrised according to \cite{Letaw}. In this way, we have
a parametrisation which is asymptotically correct at high energies
and which also fits the subthreshold data. However, we are unable to
assess the accuracy of this treatment for the problem at hand, believing
it to be at the 50 \% level, but acknowledging the need for improved
data and theoretical modeling of this seldom discussed problem.

\subsection{Primary proton flux}

The last step to make before implementing Eq.~(\ref{eq:source}) is to determine
the primary proton flux $\Phi_p(\vec{x},E^{\prime})$. The formalism introduced
in Section~\ref{sec:prop} is suitable for this purpose once we specify
the source function for primary protons. It is generally believed that
supernova remnants are the main sources of cosmic rays. Nevertheless
the gradient of $\Phi_p$ as a function of the distance from the galactic centre
which is obtained from the observed distribution of supernovae or the related
pulsar distribution is not consistent with models for gamma-ray emission
(see \cite{StrongMoskalenko}, and references therein). We take advantage
of the phenomenological approach of  \cite{StrongMoskalenko}, where
 a generic form for the radial distribution of cosmic-ray sources
was considered
and its parameters fitted to EGRET gamma-ray data (Eq.~(6) and
Fig.~12
in~\cite{StrongMoskalenko}). We therefore assume that the primary proton
source, in cylindrical coordinates, is of the form:
\begin{equation}
Q_{p}(E,\vec{x}) = \tilde{q}(E)\, q(\vec{x}) = \tilde{q}(E)\,
\left( \frac{r}{r_0}\right)^{0.5} \exp \left( - \frac{r-r_0}{r_0}\right)
\theta(h_g - |z|)
\end{equation}
where $r_0 =8.5$ kpc is our galactocentric distance, and we have assumed that
the energy spectrum of emitted protons is the same everywhere in the
Galaxy. The function $\tilde{q}(E)$, which we may interpret as a normalization factor,
can be rewritten, after propagation, in terms of the local proton flux
$\Phi_p(r_0,E)$ which has been measured in several experiments. It was argued
in the past that the spread among different experimental determinations of
$\Phi_p(r_0,E)$ introduces one of the main factors of uncertainty in the
prediction for the secondary antiproton flux (see for
instance~\cite{gaisserschaefer}). The recent measurements by the
{\sc Imax}~(\cite{imaxpro}) and {\sc Caprice} (\cite{capricepro}) 
collaborations are in better agreement with each other. In \cite{bottinolast},
a fit of the data of these
two experiments with a single power law in energy or rigidity was made, using
the force-field method to take solar modulation into account. The fits
in rigidity (see also~\cite{capricepro}) show a steeper fall-off than those
in energy; however,  this may not be the case if a break in the spectrum
at low rigidities is assumed instead~(\cite{ormespro}). As in the next Section
the background at high energies will be important, we
prefer to conservatively consider fits with a power law in energy. From
Eq.~(1) and Table~1 in~\cite{bottinolast},
\begin{equation}
\Phi_p(r_0,E) = A \frac{\sqrt{E^2-{m_p}^2}}{E}
\,\left(\frac{E}{1\,\rm{GeV}}\right)^{-\alpha}
\end{equation}
with $A = 12300 \pm 3000$ and $\alpha = 2.67 \pm 0.06$ for the {\sc Imax}
data, and $A = 19600 \pm 3000$ and $\alpha = 2.85 \pm 0.04$ for the 
{\sc Caprice} data.

\subsection{Interstellar secondary antiproton flux}

We are now ready to give the formula for the interstellar secondary
antiproton flux at our galactocentric distance. We find:
\begin{eqnarray}
\Phi_{\bar{p}}(r_0,E) & = & \frac{1}{4\,\pi} v_{\bar{p}}(E)
\,N_{\bar{p}}(r_0,E) =  \nonumber \\
& = &
2\, v_{\bar{p}}(E) \, \sum_{s=1}^{\infty} \;
J_0 \left(\nu_s^0 \frac{r_0}{R_h}\right)\, M_{\bar{p}}(s,E)\, I_{R}(s)
\int\limits_{E_{\rm thresh}}^{\infty} dE^{\prime}
\frac{\frac{d\sigma_i}{dE}(E,E^{\prime})\,n^i \; \Phi_p(r_0,E^{\prime})
\; M_p(s,E^{\prime})}
{\sum_{s^{\prime}=1}^{\infty} \;
J_0 \left(\nu_{s^{\prime}}^0 \frac{r_0}{R_h}\right)
M_p(s^{\prime},E^{\prime}) \, I_{R}(s^{\prime})}
\label{eq:sec}
\end {eqnarray}
where the repeated index $i$ stands for the sum over hydrogen and helium, and
we have introduced the notation:
\begin{equation}
I_{R}(s) = \frac{2}{{R_h}^2\,{J_1}^2(\nu_s^0)}
\int\limits_0^{R_h} dr^{\prime} \;r^{\prime}
J_0 \left(\nu_s^0 \frac{r^{\prime}}{R_h}\right) q(r^{\prime})
\end{equation}
and $M_p(s,E)=M^*(s,k=0)$ setting the total inelastic cross section
appropriate for $pp$ collisions 
$\sigma^{\rm in}_{cr\,p}=\sigma^{\rm in}_{p\,p}$,
while $M_{\bar{p}}(s,E)=M^*(s,k=0)$ with
$\sigma^{\rm in}_{cr\,p}=\sigma^{\rm in}_{\bar{p}\,p}$.
The parametrizations for both of these cross sections were given 
in~\cite{tanng2}:
\begin{equation}
  \sigma^{\rm in}_{p\,p}(T) = \left\{ \begin{array}{lcl}
    \frac{32.2 \cdot \left[ 1 + 0.0273 \ln(E/200) \right]}
    {\left[ 1 + 0.00262 \cdot 
     T^{-\left( 17.9 + 13.8 \ln T +4.41 \ln^2 T \right)} \right]} \; \rm{mb}  
    & , &  0.3 \leq T < 3  \;\rm{GeV} \\
    32.2 \cdot \left[ 1 + 0.0273 \ln(E/200) \right] \; \rm{mb} 
    & , &  3 \leq T < 200 \;\rm{GeV} \\
    32.2 \cdot \left[ 1 + 0.0273 \ln(E/200) 
    + 0.01 \left(\ln(E/200)\right)^2 \right] \;\rm{mb} 
    & , & T \geq 200\;\rm{GeV} \\
    \end{array} \right.
  \label{eq:totinp}
\end{equation}
and  
\begin{equation}
  \sigma^{\rm in}_{\bar{p}\,p}(T) = 24.7 \cdot
  \left( 1 + 0.584 \, T^{-0.115} + 0.856 \, T^{-0.566} \right) \; \rm{mb}
  \quad , \quad T \geq 0.05\;\rm{GeV}
  \label{eq:totin}
\end{equation}
where $T$ and $E$ are in units of GeV. To derive Eq.~(\ref{eq:sec}) 
we have assumed that for $z < h_g$ the approximation 
$\Phi_p(z) \simeq \Phi_p(z=0)$  is valid.
This is generally a very good approximation, as
for most choices of the parameters in the propagation model $\Phi_p$ is nearly
constant in the disk and rapidly decreasing in the halo (see for instance
Fig.~3.10 in~\cite{Berezinskii}). Only in extreme cases can $\Phi_p(z=h_g)$
be $10 \%$ lower than $\Phi_p(z=0)$ and the correction to the result
in Eq.~(\ref{eq:sec}), always below few per cent, 
can be obtained keeping track of the full $z$ dependence
in $J_p$ (use Eqs.~(\ref{eq:zdep}) and~(\ref{eq:halodiff}), all numerical
integrals in Eq.~(\ref{eq:int}) can still be performed analytically; the
result follows easily).

In \cite{bottinolast}, it was suggested that it is a good approximation
to assume that the energy spectrum for the protons {\it after propagation}
is roughly independent of location in the Galaxy. This hypothesis simplifies
the computation (for us, it is especially needed to compute numerically the
tertiary contribution described below); Eq.~(\ref{eq:sec}) reduces to:
\begin{equation}
\Phi_{\bar{p}}(r_0,E) =
\frac{2\, v_{\bar{p}}(E)\;\int\limits_{E_{\rm thresh}}^{\infty} dE^{\prime}
\frac{d\sigma_i}{dE}(E,E^{\prime})\,n^i \; \Phi_p(r_0,E^{\prime})}
{\sum_{s^{\prime}=1}^{\infty} \;
J_0 \left(\nu_{s^{\prime}}^0 \frac{r_0}{R_h}\right)
M_p(s^{\prime},\hat{E}) \, I_{R}(s^{\prime}) }
 \, \sum_{s=1}^{\infty} \;
J_0 \left(\nu_s^0 \frac{r_0}{R_h}\right)\, M_{\bar{p}}(s,E)
\, M_p(s,\hat{E}) \, I_{R}(s)
\label{eq:secsim}
\end{equation}
where $\hat{E}$ is an arbitrary normalization energy.
As there are some indications that the energy spectrum may indeed be steeper
far away from the sources, because of the energy dependence in the
propagation coefficient~(\cite{mori}),
we compare in one case Eq.~(\ref{eq:secsim}) against
Eq.~(\ref{eq:sec}) to check if the simplification  in any
way changes the result.
For the set of parameters as in example in Fig.\,\ref{fig:back}, which we 
will soon discuss, we find that Eq.~(\ref{eq:secsim})
gives a very slight overestimate of Eq.~(\ref{eq:sec}), below 3\% for
interstellar antiproton kinetic energies up to 1~GeV, a maximal 5.5\%
overestimate at 3~GeV, while for higher energies the difference decreases
again and is below 4\% at 50~GeV
(we remark however that we have not tuned our propagation model
to reproduce the effect in~\cite{mori} so we cannot claim that this effect
is not relevant).

\subsection{Tertiary antiprotons}

In Eq.~(\ref{eq:diff}) we have not introduced any energy-changing term.
We now include energy losses for secondary antiprotons due to scattering
processes during their propagation in the Galaxy. The main effect is due to
non-annihilation inelastic interactions of antiprotons with interstellar
protons, giving lower energy antiprotons in the final state.
Actually, the energy distribution after the non-annihilation interaction is
not well known as there are no direct measurements; the usual
assumption~(\cite{tanng2}) is that the distribution is similar to the final
state proton in $pp$ inelastic (non-diffractive) interactions, i.e.,
a rather flat distribution in kinetic
energy between zero and the kinetic energy of the incident antiproton.
One may think that elastic scattering processes are relevant as well but
available data show that the cross section is dominated by the forward peak
with very small energy transfer (\cite{eisenhandler}; \cite{bruckner}), and 
hence with a marginal net effect from our point of view.
We include therefore only non-annihilation processes considering a
`tertiary' source function generated by inelastically scattered secondary
antiprotons
in the form:
\begin{equation}
Q_{\bar{p}}^{\rm tert}(\vec{x},E) = 4\,\pi\,n^H(\vec{x})\;\left[
\int\limits_{E}^{\infty}
\frac{\sigma_{\bar{p}p}^{\rm non-ann}(E^{\prime})}{T^{\prime}}
I_{\bar{p}}(\vec{x},E^{\prime})dE^{\prime}\;-\;
\sigma_{\bar{p}p}^{\rm non-ann}(E) I_{\bar{p}}(\vec{x},E) \right]\;
\label{eq:tertso}
\end{equation}
where $\sigma_{\bar{p}p}^{\rm non-ann}$ is obtained as the difference between 
the total inelastic cross section Eq.~(\ref{eq:totin}) and the inelastic
annihilation cross section:
\begin{equation}
  \sigma^{\rm ann}_{\bar{p}p}(T) = \left\{ \begin{array}{lcl}
  661 \cdot \left( 1 + 0.0115 \, T^{-0.774}
    - 0.948 \, T^{0.0151} \right) \; \rm{mb}
   & , & T < 15.5 \;\rm{GeV} \\
  36 \, T^{-0.5} \;\rm{mb} 
    & , &  T \geq 15.5 \;\rm{GeV} \\
   \end{array} \right.
  \label{eq:inann}
\end{equation}
where for lower energies we have used the parametrization in~\cite{tanng2},
while in the high energy range we apply the approximation given 
in~\cite{protheroe81}. Both this parametrization and those needed
above have been checked against a compilation of more recent data
(\cite{pdg}).
The second term in Eq.~(\ref{eq:tertso}) takes into account antiprotons
which are depleted from the energy $E$ and which we propagate as a negative
flux; it actually gives an effect that is less than few per cent at any energy
and is not needed in our formalism.
As was done for Eq.~(\ref{eq:source}), 
it is straightforward to write a Fourier-Bessel
expansion for $Q_{\bar{p}}^{\rm tert}$ and then compute
$\Phi_{\bar{p}}^{\rm tert}$ which has to be summed to $\Phi_{\bar{p}}$ to 
get the final expression for the background interstellar antiproton flux.

\subsection{Numerical results}

\begin{figure}[t]
 \centerline{\epsfig{figure=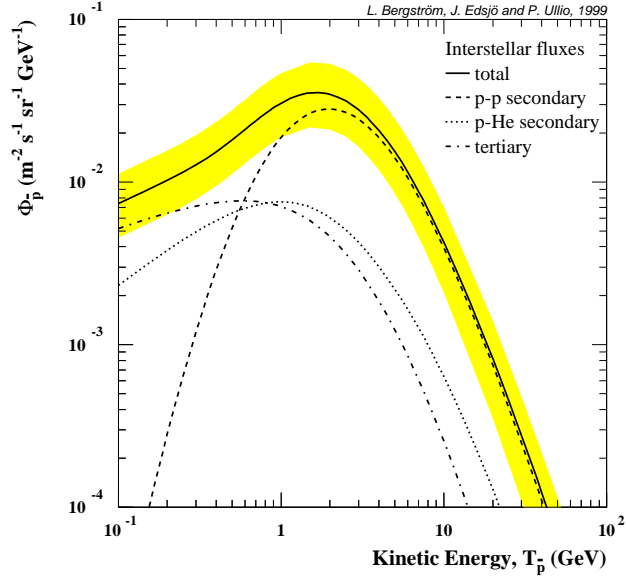,width=0.49\textwidth}}
 \caption[]{The interstellar antipron flux and the contribution from 
 secondary and tertiary antiprotons.  The uncertainty due to the 
 parametrization of the primary proton spectrum is also given as the 
 shaded band.  The solid line corresponds to the same set of 
 parameters as in Fig.~\protect\ref{fig:back}.}
 \label{fig:interst}
\end{figure}

\begin{figure}[t]
 \centerline{\epsfig{figure=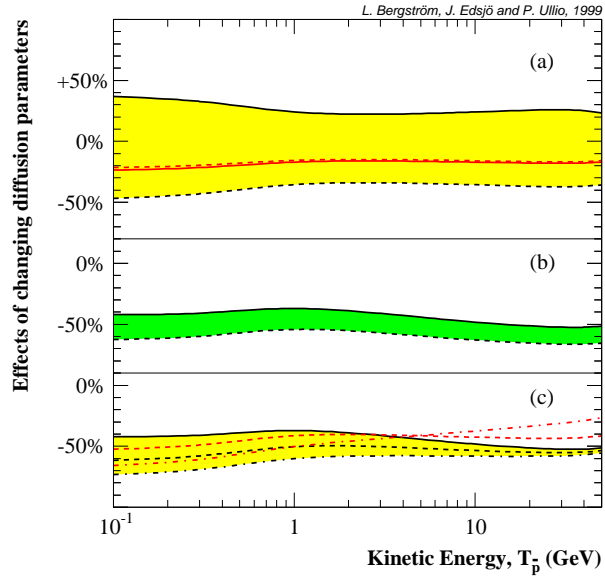,width=0.49\textwidth}}
 \caption[]{The effects of changing the diffusion parameters are
 shown. In (a) the thin halo scenario, in (b) the thick halo scenario,
 while in (c) a convective halo is considered. See the text for further
 details.}
 \label{fig:uncer}
\end{figure}

Coming to the actual numerical predictions for the background flux of 
antiprotons, we base our choice of parameters in the propagation model 
on previous work in which diffusion models analogous to the one 
described in Section~\ref{sec:prop} were used to fit data on 
cosmic-ray nuclei, such as ratios of secondaries to primaries and of 
radioactive nuclei to their stable counterparts.  Actually slightly 
discrepant results are present in the literature, partially reflecting 
the fact that it is not easy to find a propagation model which is 
consistent with the whole set of existing data.  We consider here and 
in the following Section, when describing the signal from neutralino 
annihilations, three different scenarios.  We only keep the following 
parameters fixed: $n_g^H = 1 \,\rm{cm}^{-3}$, $n_h^H = 0$, $h_g = 0.1 
\,\rm{kpc}$ and $R_h = 20 \,\rm{kpc}$.  The first three are the 
standard values which are inferred from direct observation.  The last 
one, which in the literature is taken sometimes as small as 15~kpc, 
and which in Strong \& Moskalenko (1998) is set equal to 30~kpc, does 
not play a major role and different choices lead to nearly equivalent 
results.

The three scenarios are:
\begin{itemize}
\item[\it a)] In the case $D_g^0 = D_h^0 = D^0$ our propagation model
  is fairly close to the one considered by Webber, Lee and Gupta
  (1992). Their conclusion is that a thin halo is preferred, with
  height $h_h \in (1.1,3.8) \,\rm{kpc}$ and $D \simeq (6 \pm 4)\cdot
  10^{27} \,\rm{cm}^2 \rm{s}^{-1}$ at the rigidity $R = 1 \,\rm{GV}$.
  The flux shown in Fig.~\ref{fig:back} is obtained in this scenario,
  setting $h_h = 3 \,\rm{kpc}$, $D^0 = 6 \cdot 10^{27} \,\rm{cm}^2
  \rm{s}^{-1}$ and $R_0 = 3 \,\rm{GV}$, choosing the proton flux at
  the Earth as the medium value in the fit of IMAX data, i.e. with $A
  = 12300$ and $\alpha = 2.67$, and taking into account solar
  modulation with the force field method with $\phi_F = 500 \,
  \rm{MV}$ as suggested by the analysis of the {\sc Bess} collaboration.
  There is no consensus in the literature on the value of the solar
  modulation parameter at solar minimum, nevertheless the spectrum
  does not change dramatically if a slightly different value for
  $\phi_F$ is assumed. For instance, $\phi_F = 400 \, \rm{MV}$ gives
  about a $7\%$ increase at the kinetic energy $T = 0.2~\rm{GeV}$ and
  about an $8\%$ increase at the maximum. For $\phi_F = 600 \, \rm{MV}$
  the effect is reversed and we find that the flux is lower by roughly
  the same percentages as in the previous case.
  
  As will become clear in the following, the background antiproton
  flux shown in Fig.~\ref{fig:back} is only an example of the
  possibility of a good fit to the data; we keep it as reference case
  to compare with. In Fig.~\ref{fig:interst}, we show for the same
  parameters the interstellar antiproton flux versus kinetic energy
  $T$, plotting also its three main components: the secondary
  antiproton flux due to $pp$ collisions, the contribution from $pHe$
  scattering processes and the tertiary component due to energy loss.
  As can be seen the first contribution is dominant at the maximum and
  at high energies, while the other two are important in the low
  energy region.
  
  We take advantage of Fig.~\ref{fig:interst} to show another feature
  that is common for all choices of the propagation parameters, the
  uncertainty due to the interstellar proton flux. The band around our
  reference antiproton flux is the envelope of the predictions
  obtained by using the uncertainty in the proton flux
  (\cite{bottinolast}): the upper bound is given choosing the fit of
  IMAX data with $A = 15300$ and $\alpha = 2.61$ (average values $+ 1
  \sigma$ and $- 1 \sigma$ respectively), while the lower bound below
  $T = 2.5\,\rm{GeV}$ is obtained from the IMAX data fit with $A =
  9300$ and $\alpha = 2.73$ and above 2.5~GeV from the CAPRICE data
  fit with $A = 16300$ and $\alpha = 2.89$ (average values $- 1
  \sigma$ and $+ 1 \sigma$ respectively, actually these two spectra
  are nearly overlapping at all energies).
  
  Coming back to the uncertainty in the choice of the propagation
  parameters in the Webber-Lee-Gupta scenario, if we now pick the
  average value for the halo height $h_h = 2\, \rm{kpc}$ and vary the
  diffusion coefficient in the suggested interval, $D^0 \simeq (3-7)
  \cdot 10^{27} \,\rm{cm}^2 \rm{s}^{-1}$ for $R_0 = 3 \,\rm{GV}$, we
  find that the the flux at intermediate energies increases by up to
  about $30 \%$ for the smallest value of the diffusion coefficient,
  while it decreases with a slightly higher percentage for the highest
  value of $D^0$. This is represented by the band in
  Fig.~\ref{fig:uncer}, region (a) (both in this Figure and in
  Fig.~\ref{fig:cpropunc} below we defined fractional differences as
  $(\Phi - \Phi_R) / \Phi_R$, with $\Phi_R$ being the reference value,
  i.e. in this case the flux shown as a solid line in
  Fig.~\ref{fig:interst}).  If we assume on the other hand that $D^0$
  and $h_h$ are linearly related, a degeneracy which may indeed not be
  resolved by available data, and fix $D^0 = 2.5 (h_h/\rm{kpc}) \cdot
  10^{27} \,\rm{cm}^2 \rm{s}^{-1}$ varying $h_h$ between 1.1 and
  3.8~kpc we get a band of very small width, below a few percent (solid
  and dashed lines at about $-20 \%$ in Fig.~\ref{fig:uncer}, region
  (a)).

\item[\it b)] In ~\cite{StrongMoskalenko}, a scenario is favoured with
  a thicker halo, with $h_h \in (4,12) \,\rm{kpc}$, in case no
  convection is assumed. Their treatment of propagation is less close
  to our model than the previous case, so the way we translate their
  typical choice of parameters into our picture is less safe but
  should give at least the right qualitative behaviour. We sketch the
  thick halo scenario taking $h_h \in (4,12) \,\rm{kpc}$, $D_h^0 = 2.5
  (h_h/\rm{kpc}) \cdot 10^{27} \,\rm{cm}^2 \rm{s}^{-1}$, $D_g^0 = 6
  \cdot 10^{27} \,\rm{cm}^2 \rm{s}^{-1}$ and $R_0 = 1 \,\rm{GV}$, a
  choice consistent with the results in \cite{GKP}, where the
  propagation model we have chosen was first considered.  Comparing to
  our reference case we find a band of $20 \%$ width around an average
  suppression of the flux of about $50 \%$, where the least severe
  suppression is given by the smallest halo considered and the maximal
  suppression corresponds to the large halo $h_h = 12 \,\rm{kpc}$
  (solid and dashed lines respectively in Fig.~\ref{fig:uncer} region
  (b)).
  
\item[\it c)] As a third scenario we allow for the presence of a
  galactic wind driving cosmic rays out of the galactic disk. Self
  consistent models for the propagation of cosmic rays in
  magnetohydrodynamic flows have been studied recently in
  detail~(\cite{zirakashvili}; \cite{ptuskin}). The much simpler
  approach we take here is intended to compare qualitatively the
  effects of the wind on the background antiproton flux and on the
  signal from neutralino annihilations. Considering again the model in
  the previous scenario with $h_h = 4 \,\rm{kpc}$ (solid curve in
  Fig.~\ref{fig:uncer} region (c)), we take as an example the case of
  $v_h = 10\,\rm{km\;}\rm{s}^{-1}$ (lower dashed line) and $v_h =
  20\,\rm{km\;}\rm{s}^{-1}$ (lower dash-dotted line).  As can be seen,
  in perfect analogy with the case of the solar wind, the galactic
  wind alters the spectrum at low energies (up to $30 \%$ in the
  example we are considering) while the effect gets smaller and
  smaller for more energetic particles. If in analogy with the
  parameter choice of Strong and Moskalenko we suppose that there is
  some scaling between $v_h$ and $D_h^0$, for instance a simple linear
  scaling $D_h^0 = 2.5 (h_h/\rm{kpc}) [(40\,\rm{km\;}\rm{s}^{-1} -
  v_h)/40\,\rm{km\;}\rm{s}^{-1}]\cdot 10^{27} \,\rm{cm}^2\; \rm{s}^{-1}$,
  we find that at intermediate energies the flux is nearly unchanged
(upper dashed and dash-dotted lines).

\end{itemize}

We will not combine the uncertainty bands we have just derived and make
a definite statement about the uncertainty on the prediction of the
cosmic antiproton background. To be able to do that on a firmer basis
we should compare the predictions of our propagation model directly
against the whole set of data on cosmic ray nuclei and this is
beyond the aim of the paper. We stress again that this Section was mainly
intended to show that the most recent data on cosmic ray antiprotons can 
be fitted by the background flux for some natural choice of the diffusion
parameters. On the other hand we find that the prediction for the
background could be lower as well, leaving room for an antiproton flux
generated by an exotic source, possibly dark matter neutralinos.

\subsection{Reacceleration}

It seems very plausible that cosmic rays are reaccelerated by a Fermi
type of acceleration by stochastic magnetic fields during
propagation. This has been treated, e.g.,\ in
Seo \& Ptuskin (1994); Heinbach \& Simon (1995)  and Simon \& Heinbach (1996).  
There is also a possibility that cosmic rays get reaccelerated by weak 
shock waves from supernova remnants (\cite{lst}).
We will here focus on the former process, usually called diffusive 
reacceleration since it can be treated as a diffusion in momentum space.

In Heinbach \& Simon (1995) it was shown that data on low energy 
cosmic rays are compatible with the predictions of models without 
diffusive reacceleration only if the mean path length variation with 
energy shows a sharp break around 1--2 GeV. In models which include 
reaccelaretion effects on the other hand, depending on reacceleration 
strength, a path length distribution that is a simple power law for 
all energies may be considered.  This is theoretically appealing since 
this from observations derived form agrees well with that expected 
from Kolmogorov turbulence.  The result in Heinbach \& Simon (1995) is 
confirmed in the analysis by Strong \& Moskalenko (1998), who conclude 
as well that the reacceleration scheme allows a more natural choice 
for the parameters in the propagation model.

Even though reacceleration implies that the average energy increases 
as the cosmic rays propagate through the galaxy, there is a general 
smearing of the injected spectrum meaning that there is also a 
`leakage' of cosmic rays from higher energies to lower.

This might be important for antiprotons where the injected spectrum 
drops below the maximum at few GeV\@.  In Simon \& Heinbach (1996) it 
was found that this leakage could substantially increase the secondary 
$\bar{p}$ spectrum below 1~GeV, with the flux at a few hundred MeV 
that hardly can be lower than 1/3 of the value at around 1~GeV. We 
cannot compare directly with our analysis as we are not applying the 
same primary proton spectrum and propagation model, but still we can 
conclude that including reacceleration in the propagation model might 
add up to the effects of the proton-nuclei interactions and of the 
tertiary component in flattening the antiproton spectrum at low 
energies, and making it even more problematic to separate an exotic 
signal from the background in this region of the antiproton spectrum.

\pagebreak
\section{Signal from neutralino annihilation}

\begin{figure}[t]
 \centerline{\epsfig{figure=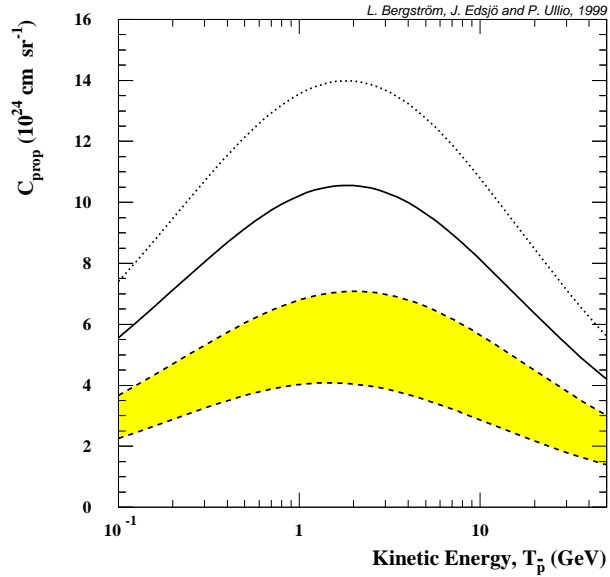,width=0.49\textwidth}}
 \caption[]{The value of $C_{\rm prop}$ for the same choice of 
 diffusion parameters as in Fig.~\protect\ref{fig:back} and an 
 isothermal sphere distribution of dark matter with $a = 3.5$~kpc 
 (solid line), compared to the value for the same diffusion parameters 
 and a Navarro et al.  profile with $a = 9$~kpc (dotted line).  The 
 band gives the range of values of $C_{\rm prop}$ in the thin halo 
 scenario (with $h_{h}=2$ kpc) as described in the text.}
 \label{fig:cprop}
\end{figure}

\subsection{General discussion}

With the source function introduced in Eq.~(\ref{sourcefkn}), the antiproton
flux from neutralino annihilations in the galactic halo is readily obtained
from the formulas derived in Section~\ref{sec:prop}. It is given by:
\begin{equation}
\Phi_{\bar{p}}(r_0,T) =  \frac{1}{4\,\pi} v_{\bar{p}}(T)
\,N_{\bar{p}}(r_0,T) = \frac{1}{4\,\pi} v_{\bar{p}}(T)
\sum_{s=1}^{\infty} \; J_0 \left(\nu_s^0 \frac{r_0}{R_h}\right)\, M_s^0(0)
\label{eq;signal}
\end {equation}
where $M_s^0(0)$ is obtained from Eq.~(\ref{eq:halodiff}) with
$\sigma^{\rm in}_{cr\,p}=\sigma^{\rm in}_{\bar{p}\,p}$ in
Eq.~(\ref{eq:halodiff2}) and $Q_s^0(z)$ in Eq.~(\ref{eq:int}) given by
\begin{equation}
Q_s^0(z) = \frac{2}{{R_h}^2\,{J_{1}}^2(\nu_s^0)}
\int\limits_0^{R_h} dr^{\prime} \;r^{\prime}
J_0 \left(\nu_s^0 \frac{r^{\prime}}{R_h}\right)
\,Q_{\bar{p}}^{\chi}(r^{\prime},z)\;\;.
\end{equation}

It is possible to separate in the expression for the signal the
part which depends on the MSSM parameter space from the terms which
are related only to the distribution of sources in the propagation region
and to the propagation model itself. We introduce the definition:
\begin{equation}
  \Phi_{\bar{p}}(r_0,T) \equiv
  (\sigma_{\rm ann}v) \sum_{f}^{}\frac{dN^{f}}{dT}B^{f}
  \left(\frac{\rho_0}{m_{\tilde{\chi}}}\right)^{2}
  \, C_{\rm prop}(T) \;.
  \label{eq:signal2}
\end{equation}
The quantity $C_{\rm prop}$, which can be obtained explicitly by comparing
Eq.~(\ref{eq;signal}) with Eq.~(\ref{eq:signal2}),
has the dimension of length divided by solid angle and is analogous to the
coefficient defined in Eq.~(46) of~\cite{bottinolast}; note however that
in Eq.~(\ref{eq:signal2}) we have factorized the value of the local halo
density $\rho_0$ rather than some reference density.

\subsection{Uncertainties related to propagation}

Having factorized out in Eq.~(\ref{eq:signal2}) the dependence of the
signal on the choice of the dark matter candidate, we analyse first
how sensitive the result is to the set of parameters which define both
the location of the sources and the propagation of the produced
antiprotons. We fix a reference configuration selecting for the
propagation model, in analogy to the analysis of the background flux,
the same parameters as in the example in Fig.~\ref{fig:back} and
Fig.~\ref{fig:interst} ($h_h = 3 \,\rm{kpc}$, $D^0 = 6 \cdot 10^{27}
\,\rm{cm}^2 \rm{s}^{-1}$ and $R_0 = 3 \,\rm{GV}$), while as a reference
dark matter density profile we choose a modified isothermal
distribution, Eq.~(\ref{eq:haloprof}) with
$(\alpha,\beta,\gamma)=(2,2,0)$, and with an intermediate value for
the length scale, $a = 3.5$~kpc.  In Fig.~\ref{fig:cprop} we plot the
value of $C_{\rm prop}$ for this reference case (solid curve) versus
the antiproton kinetic energy.  $C_{\rm prop}$ is increasing in the
low energy range as it contains the kinematic factor $v_{\bar{p}}(T)$
and at the same time we have assumed that the diffusion coefficient is
roughly constant at low rigidities (see Eq.~(\ref{eq:diffco})). The
coefficient $C_{\rm prop}$ then reaches a maximum at about $T \simeq
2$~GeV, while at higher energies it decreases as a consequence of the
0.6 power law increase in the diffusion coefficient.

\begin{figure}[t]
 \centerline{\epsfig{figure=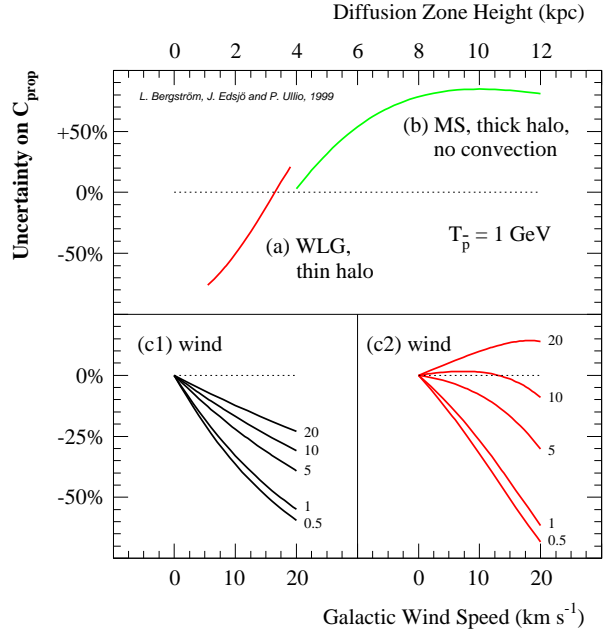,width=0.49\textwidth}}
 \caption[]{The changes of $C_{\rm prop}$ when the diffusion zone
 height is changed within (a) the thin halo scenario derived from
 Webber, Lee and Gupta model and (b) the thick halo scenario inspired by
 the Moskalenko and Strong analysis. In (c) the effects of a convective wind
 are sketched: in (c1) the diffusion coefficient is kept fixed while
 in (c2) it is linearly scaled with the galactic wind.}
 \label{fig:cpropunc}
\end{figure}

First we analyse how the result changes if as the dark matter density 
distribution we consider the Navarro et al.\ profile (\cite{navarro}) 
which is singular towards the galactic centre, Eq.~(\ref{eq:haloprof}) 
with $(\alpha,\beta,\gamma)=(1,2,1)$.  Choosing $a = 9$~kpc, we obtain 
(see the dotted line in Fig.~\ref{fig:cprop}) roughly a 33\% increase 
in $C_{\rm prop}$ (and therefore in the signal from neutralino 
annihilations) at any T, while the more cuspy profile with $a = 
3.5$~kpc gives a result which is more than twice the reference value.  
This is rather surprising because even though the singularity in the 
profile induces a sharp enhancement in the neutralino number density 
and therefore in the strength of the source, this cusp is at the 
galactic centre rather far away from the solar system.  It is commonly 
believed that in the diffusion regime the local sources are the most 
relevant, but at least in the propagation model we are considering, 
this is not true: for the Navarro et al.\  profile with $a = 9$~kpc, 
23\% of the signal is given by sources contained in a spherical region 
of 1 kpc around the galactic centre; the percentage increases to 42\% 
for $a = 3.5$~kpc while it is as low as 1\% for the isothermal sphere 
profile.  We conclude that non-local sources may give a significant 
contribution provided their strength is much enhanced with respect to 
the local ones.  This effect should be considered in more detail when 
considering a clumpy scenario~(\cite{pieroclumpy}) for which it may be 
even more relevant.

Fig.~\ref{fig:cprop} contains another piece of information. Going back to
the case where the dark matter density profile is described by an isothermal
sphere, we have varied the parameters which define the propagation
model, going through the same three scenarios described when discussing the
background. The band in the figure is given by fixing
$h_h = 2\, \rm{kpc}$ and varying the diffusion coefficient in the interval
$D^0 \simeq (3-7) \cdot 10^{27} \,\rm{cm}^2 \rm{s}^{-1}$, and corresponds
to the band shown in part (a) of Fig.~\ref{fig:uncer} (again the highest
value of $D^0$ gives the lowest value for the flux). Unlike the latter,
the band in Fig.~\ref{fig:cprop} does not overlap the reference value 
(solid line): for the signal from neutralino annihilations
the decrease in the height of the propagation zone from 3~kpc to 2~kpc is
not compensated by the decrease in the central value for the diffusion
coefficient. This gives a first hint on how sensitive the dark matter signal
is to the choice of the value of the height of the diffusion zone.

The same effect is studied in the upper part of Fig.~\ref{fig:cpropunc},
fixing the kinetic energy to $T = 1$~GeV, varying $h_h$ and linearly
relating $D^0$ to $h_h$, as introduced in the previous Section.
According to the Webber-Lee-Gupta scenario, $h_h$ is constrained to be
between 1.1 and 3.8~kpc: the degeneracy we found for the background flux
(the nearly overlapping solid and dashed line in part (a) of 
Fig.~\ref{fig:uncer}) is completely removed for the signal from neutralino 
annihilations, going from an 80\% suppression to a 20\% increase compared 
to the reference value, for minimal and maximal $h_h$ respectively. 
Changing the diffusion zone height modifies the number of sources which 
contribute to the flux, as sources which are outside the diffusion box are 
not included in the model. Therefore it is not  
surprising that a very thin diffusion zone gives a suppressed signal 
while larger values  for $h_h$ enhance it.

In the same way, in the thick halo scenario without convection, inferred
from the analysis of Strong and Moskalenko, the suppression band found for 
the background flux (part (b) of Fig.~\ref{fig:uncer}) becomes a much wider 
band for which the signal flux is increased instead (solid line in part (b) of
Fig.~\ref{fig:cpropunc}).
The enhancement with the diffusion zone height is flattened out at high 
values of $h_h$ as the new sources we include are further and further 
away from the observer and moreover the density profile falls at
large galactocentric distances.

While the effects we have considered so far give roughly the same 
result for any value of the antiproton kinetic energy, the effect of 
convection in the z direction is clearly energy dependent.  In part 
(c1) of Fig.~\ref{fig:cpropunc} we plot $C_{\rm prop}$ as a function 
of the galactic wind speed, for $h_h = 4 \,\rm{kpc}$, $D_h^0 = 10^{28} 
\,\rm{cm}^2 \rm{s}^{-1}$, $D_g^0 = 6 \cdot 10^{27} \,\rm{cm}^2 
\rm{s}^{-1}$ and $R_0 = 1 \,\rm{GV}$ (we consider here as reference 
value the one obtained with this set of parameters and $u_h = 0$).  In 
part (c2), as we did for the background, $D_h^0$ is taken to be 
related linearly to the value of the wind speed.  In both parts five 
different kinetic energies (given in the Figure in GeV) have been 
considered.  As can be seen going back to part (c) of 
Fig.~\ref{fig:uncer}, the effect of convection is greater on the 
signal from neutralino annihilation than on the background flux.  
Especially at low energies, it is not well compensated by the linear 
scaling of the diffusion coefficient.

\begin{figure}[tb]
\centerline{\epsfig{file=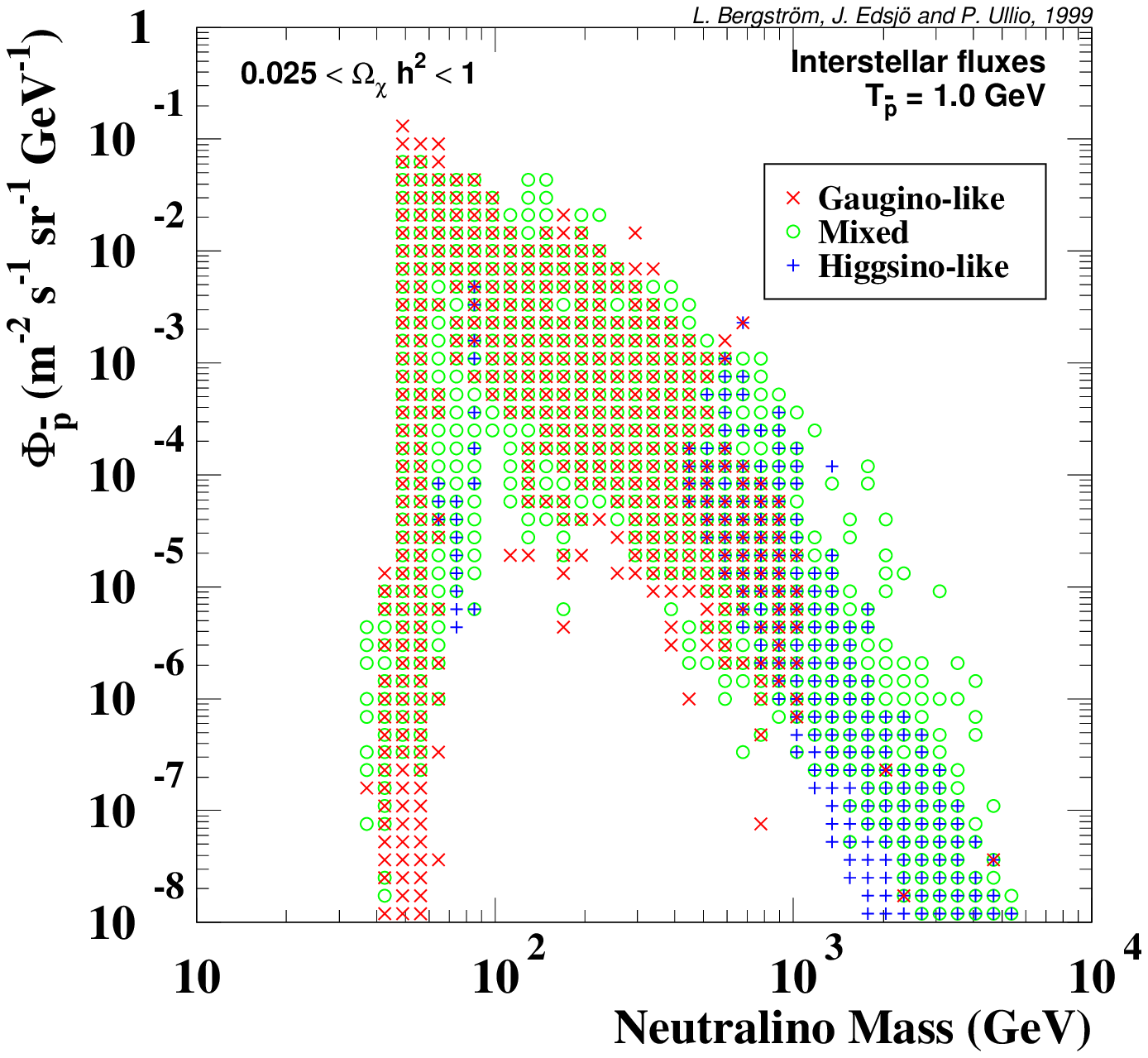,width=0.49\textwidth}
\epsfig{file=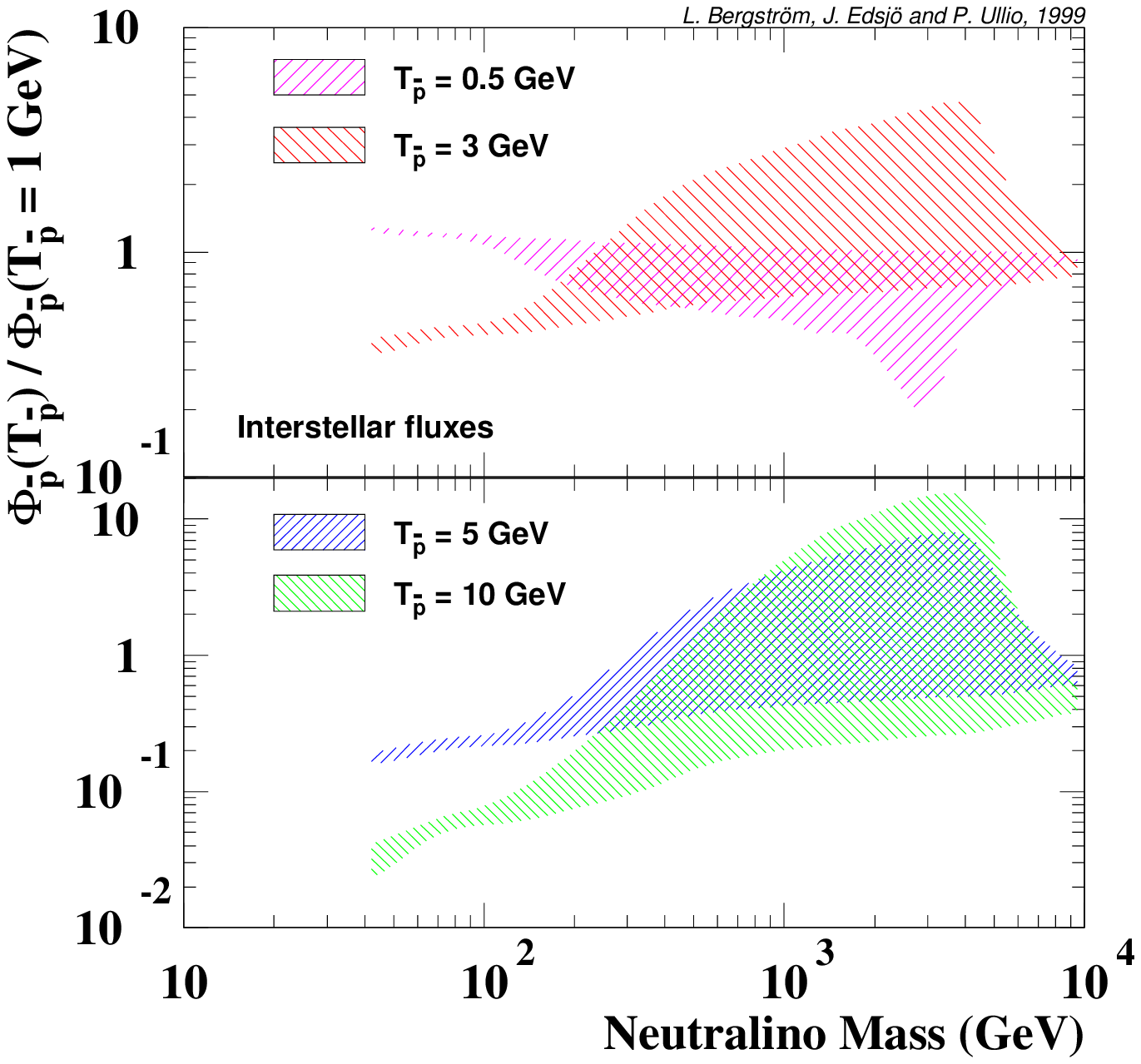,width=0.49\textwidth}} \caption{In (a) 
the interstellar antiproton flux at 1 GeV is shown and in (b) the 
ratio of the flux at different kinetic energies to that at 1 GeV is 
shown.  To make the figure clearer (and avoid showing artifacts of 
sampling frequency) the figure is binned.  We also show with different 
symbols in which bins there are models which are gaugino-like, mixed 
and Higgsino-like.}
\label{fig:pbismx}
\end{figure}

For the background flux the effect for introducing a galactic wind is to
drive antiprotons more quickly from the disk, where they are generated,
to the border of the diffusion zone where they are lost. This effect can be
balanced by lowering the diffusion coefficient, that is, assuming that
diffusion takes place less efficiently. The sources of the signal are
on the other hand distributed over the whole diffusion box; setting with the
galactic wind a preferred direction of propagation lowers the probability
for an antiproton generated relatively far away from the disk to reach our
location. The effect is not compensated by the rescaling of $D_h^0$, at
least not for the rescaling needed for cosmic ray species generated in
the disk.

The last check we perform regards the role played by the antiprotons which are 
produced outside the propagation region. Our solution to the diffusion 
equation has been derived under the hypothesis that the number density 
of the considered cosmic ray species is zero at the boundary of the 
diffusion zone. This is not strictly true for the signal from neutralino 
annihilations. One possibility of verifying what kind of corrections might
be needed in this case is to compare the antiproton flux leaving the diffusion
zone with the flux injected by external sources. We restrict the analysis
to exchanges at the boundary $z = \pm h_h$, as the effect is much suppressed
in the radial direction, being generally $R_h >> h_h$.

The outgoing flux can be computed by keeping track of the full dependence on 
$z$ of the number density, as this flux is related to the gradient of the
number density at the boundaries; one derives a rather lengthy expression
which we do not reproduce here but which follows in a straightforward way.
For the injected flux we use the very simple picture of propagation in free 
space, summing contributions over the line of sight. For the 
modified isothermal sphere profile and a diffusion zone height of 3~kpc 
we find that the ingoing flux is about one third of the outgoing flux 
for very small radial coordinates, while they become roughly equal at 
$r = 8$~kpc and at larger radii the injected flux becomes prevailing; 
the total number of antiprotons per second that penetrate the diffusion 
box is about 70\% of those that leave it. 

As can be understood, this fraction is smaller if we consider instead 
the Navarro et al.\,profile or if we pick a higher value for the 
diffusion zone height, and might turn into a rather large number for 
very thin haloes.  To give a precise numerical estimate for the effect 
one should add in the propagation model a third zone above $|z| = 
h_h$. However, taking into account all the other uncertainties that 
enter in the prediction for the signal, we do not consider this 
worthwhile at present.  We believe that a safe assumption is that the 
signal from neutralino annihilations has not been underestimated due 
to this effect by more that a factor of 2 in the most extreme cases.

\subsection{Antiprotons from specific MSSM models}

\begin{figure}[tb]
\centerline{\epsfig{file=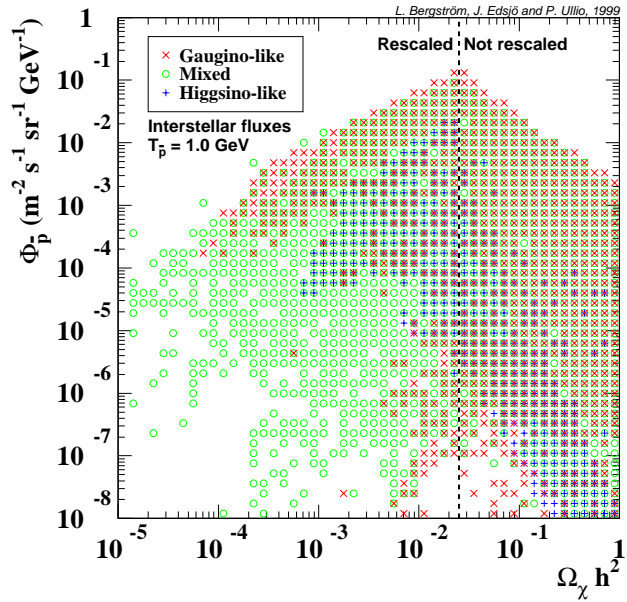,width=0.49\textwidth}}
\caption{The interstellar antiproton flux at 1 GeV versus
the relic density. This is the only figure where models having 
$\Omega_{\chi}h^2<0.025$ are shown.}
\label{fig:pbisoh2}
\end{figure}

The antiproton spectrum from neutralino annihilation have been
calculated for all the different MSSM models given in
Table~\ref{tab:scans}. In Figs.~\ref{fig:pbismx}--\ref{fig:nicemodel}
we show our main results.  We use our canonical parameters for 
propagation and the isothermal sphere model for the halo profile.

\subsubsection{Interstellar fluxes}

In Fig.~\ref{fig:pbismx} (a) we show the predicted interstellar
antiproton flux at a kinetic energy of 1 GeV (i.e., without
corrections for solar modulation) versus the neutralino mass.
We clearly see the trend that the flux goes down
with the mass of the neutralino. The reason for this is that the
number density of neutralinos goes down as the mass increases, for a
given interval of the dark matter mass density.  Since
$n_\chi=\rho_\chi/m_\chi$, and the annihilation rate scales as
$n_\chi^2$ the suppression increases rapidly with mass.
At the lower mass end, the present accelerator limits preclude a
neutralino in the MSSM below a few tens of GeV. The low-flux models 
at low masses will be discussed in connection to 
Fig.~\ref{fig:pbisoh2} below.

The points in the figure are coded with different symbols for
different composition of the neutralino. We define models with
$0< Z_g< 0.01$ as Higgsino-like, $0.01 < Z_g < 0.99$ as mixed and
$0.99< Z_g< 1$ as gaugino-like. As can be seen, most of the models
with high rates are either gaugino-like or mixed, except at masses
greater than several hundred GeV, where also Higgsino-like models can
be important. (In fact, there is also a small mass window around 80
GeV where Higgsinos may be relevant.)

In Fig.~\ref{fig:pbismx} (b), we show the ratio of the interstellar flux
at 0.5, 3, 5, and 10 GeV to the flux at 1 GeV displayed in (a) for the
same set of models (but without coding the composition). It is seen
that for the higher mass models, it can be more advantageous to study
the flux at higher kinetic energies.

\begin{figure}[tb]
\centerline{
\epsfig{file=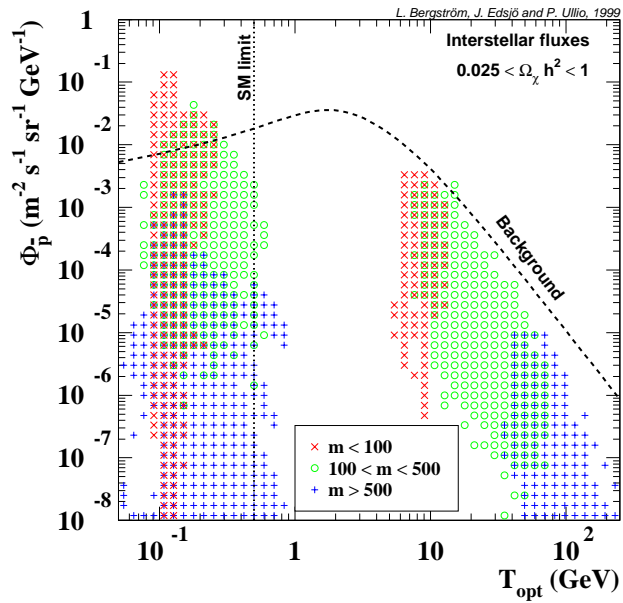,width=0.49\textwidth}}
\caption{The flux of antiprotons from neutralino annihilation at 
  the optimal kinetic energy, $T_{\rm opt}$, versus $T_{\rm
  opt}$. $T_{\rm opt}$ is defined as the energy at which $\Phi_{\rm
  signal}/\Phi_{\rm background}$ is highest and if
  the spectrum has more than one optimum, the highest two have been
  included in the plot. The models have been coded according to the
  neutralino mass in GeV\@.}
\label{fig:pbtopt}
\end{figure}

In Fig.~\ref{fig:pbisoh2} we show the same fluxes as in
Fig.~\ref{fig:pbismx} (a), but versus the relic density, $\Omega_\chi
h^2$. There is a very clear trend that the highest flux is obtained
when $\Omega_\chi h^2$ is close to the lowest acceptable relic
density. The reason for this is that if the annihilation cross section
is increased, the flux of antiprotons increases, but since the relic
density $\Omega_\chi h^2$ is approximately inversely proportional to
the annihilation cross section, $\Omega_\chi h^2$ decreases and hence
the strong correlation. The correlation is not perfect however, since
it is the thermally averaged cross section at a temperature of about
$m_\chi/20$ that determines $\Omega_\chi h^2$, whereas the
annihilation in the halo to a very good approximation occurs at rest
(the speeds are typically $\sim 0.001c$)\@.

This also exlplains some features in Fig.~\ref{fig:pbismx} (a).  In
the mass range between 40 and 60 GeV there exist models which give
exceedingly small rates, but also some which give the highest of all
rates. This large spread reflects peculiarities near the $Z^0$ (and
neutral Higgs) resonances and the $W^+W^-$ threshold.  For the
low-flux models around 40--60 GeV, the annihilation cross section at
rest is very small, but either a resonance or threshold can be reached
through thermal motion in the early Universe and the relic density is
reduced to our selected range $0.025<\Omega_\chi h^2<1$\@. As shown by
Chen \& Kamionkowski (1998), three-body final states can be important
(not too far) below the $W^+W^-$ and the $t\bar{t}$ thresholds and
this could enhance the signal for these low-rate models.

We also have some models around 130 GeV that give high fluxes. In this
case it is the other way around, the masses are just so that we are on
the $H_3^0$-resonance for the non-relativistic speeds in the halo, but
the thermal average in the early Universe gives a lower annihilation
cross section and hence the relic density is increased to our desired
range $0.025<\Omega_\chi h^2<1$\@.  The behaviour at 130 GeV is just
accidental, it could happen at any important resonance. In fact, we
found only one high-flux model around 130 GeV in our `normal' scans,
and performed a small scan varying the parameters slightly around this
model. The relic density was essentially unchanged, but the antiproton
flux showed large variations depending on if we were below, on or
above the resonance.

In Fig.\,~\ref{fig:pbisoh2} we also show models with a value of
$\Omega h^2$ lower than our required limit of 0.025. In principle, one
could accept these models at the expense of introducing other
components of dark matter. To be consistent, one should then rescale
the local dark matter density in the form of neutralinos by some
unknown factor. In the lack of better procedures, one usually employs
a linear rescaling $\rho_\chi=(\Omega_\chi h^2/0.025)\cdot
\rho_{\rm DM}$. Since the annihilation rate is quadratic in the number
density, this rescaling factor enters squared in the predicted $\bar
p$ rate, something which is clearly visible in
Fig.\,~\ref{fig:pbisoh2}.

We are now interested in finding out if there are any special features
of the antiproton spectra from neutralino annihilation which
distinguish these spectra from the background. We have already
mentioned that the window at low energies may not be as good as
previously thought. We will here investigate other features and energy
regions of the spectrum to see if there are good signatures of a
neutralino contribution to the flux.

One thing that might differ is the slope at different energies. The
background is expected to have a rising trend at low energies,
reaching a maximum between 1 and 2 GeV (see Fig.\,~\ref{fig:interst})
and a slope of around $-3$ at high energies. 
On the other hand, the high-rate models tend to be decreasing
at 1 GeV (see Fig.~\ref{fig:nicemodel} (a)). 
This may cause a shift of the maximum of the summed spectrum
(signal plus background) to a lower energy, which is a possible
signature. 

We next investigate if there is an optimal energy at which $\Phi_{\rm
signal} / \Phi_{\rm background}$ has a maximum (for this purpose we 
will use the reference background given in Fig.~\ref{fig:interst}). In
Fig.~\ref{fig:pbtopt} we show the flux at these optimal energies,
$T_{\rm opt}$, versus $T_{\rm opt}$. We now see that we have two
classes of models. One class which have highest signal to noise below
0.5 GeV (i.e.\ inaccessible in the solar system due to the solar
modulation) and one which have highest signal to noise at 10--30 GeV.
For this first class of models, we note that there exist a proposal of
an extra-solar space probe (\cite{spaceprobe}) which would avoid the
solar modulation problem and is thus an attractive possibility for
this field. However, these models have high rates in the range 0.5--1
GeV as well, even though it would be even more advantagous to go to
lower energies.  The second class of models are much less affected by
solar modulation and also give reasonably high fluxes. In
Fig.~\ref{fig:nicemodel} (a) we show some examples of spectra. All of
these have optimal energies in the low-energy region, but e.g.\
spectrum 3 has an optimum at high energy as well.

\subsubsection{Solar modulated fluxes}

\begin{figure}[tb]
\centerline{\epsfig{file=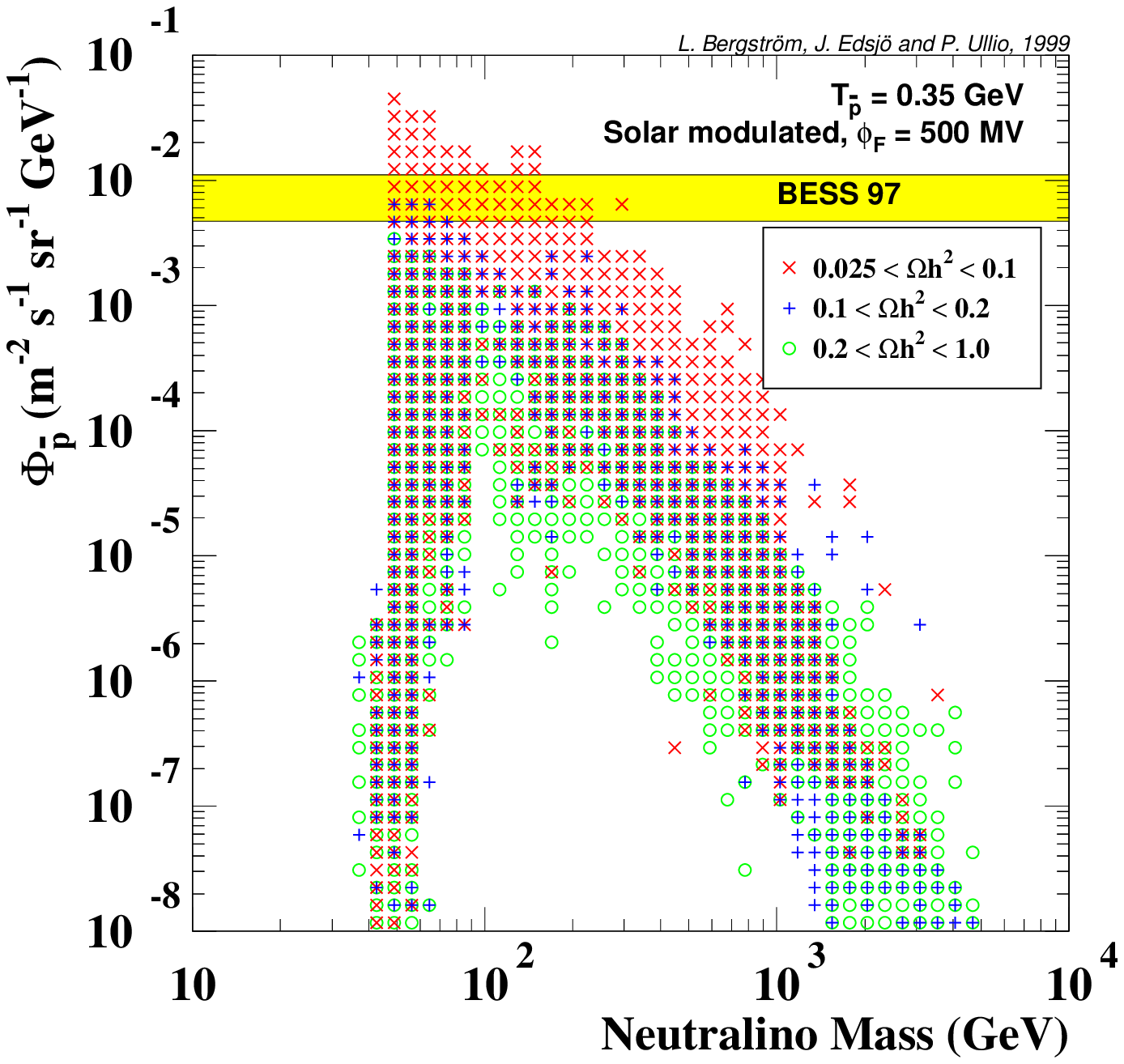,width=0.49\textwidth}
\epsfig{file=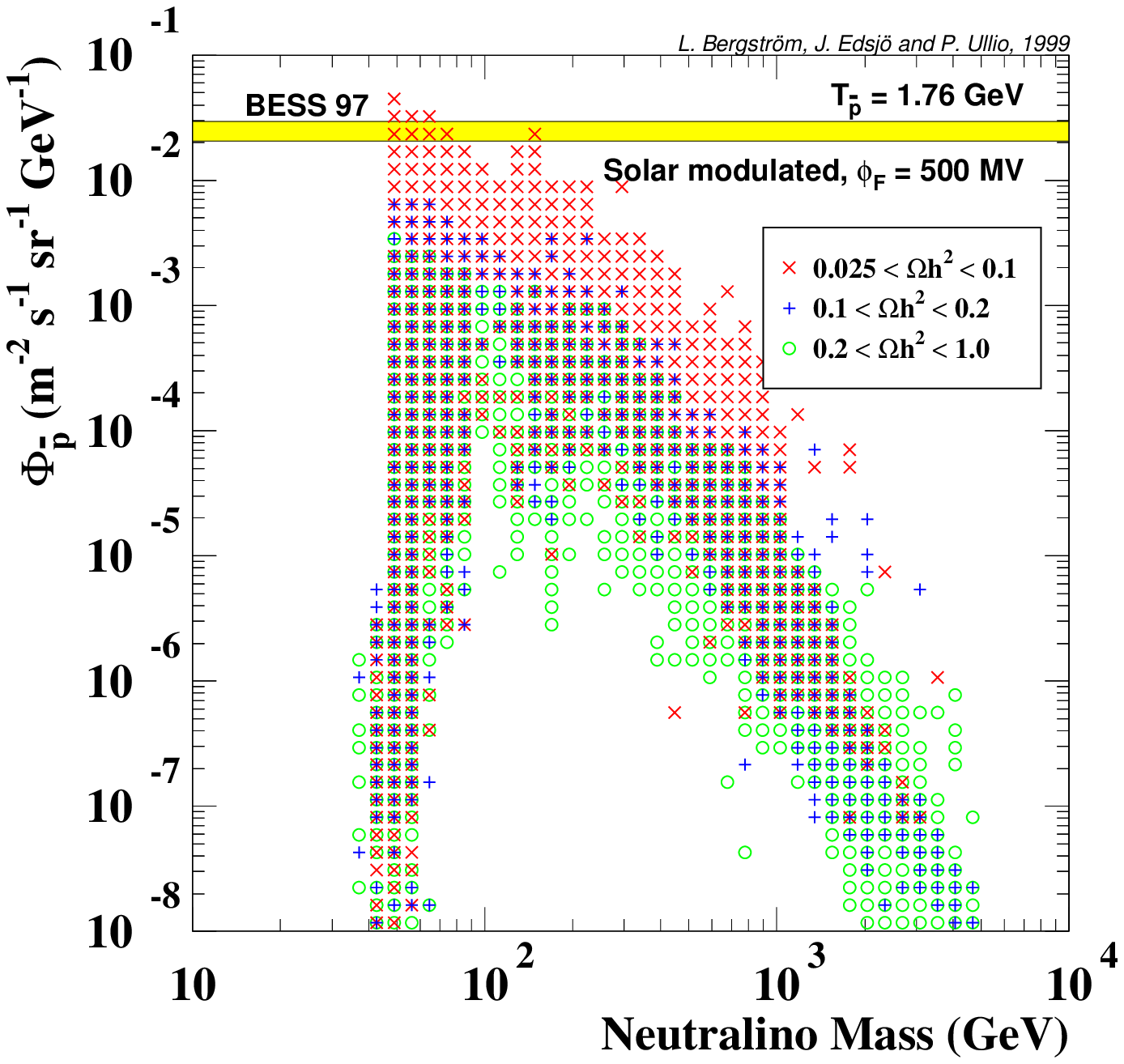,width=0.49\textwidth}}
\caption{The solar modulated antiproton fluxes at (a) 0.35 GeV and (b)
1.76 GeV compared with {\sc Bess} 97. The models have been coded according
to their relic density, $\Omega_\chi h^2$.}
\label{fig:pbsmmx}
\end{figure}

We now turn to the solar modulated fluxes and will compare with the {\sc
Bess} 97 measurements, which we recall are in very good agreement with
our estimate for the background flux.  We will compare the fluxes in
two of the {\sc Bess} energy bins, the one at 0.35 GeV and the one
where the measured flux is the highest. In Fig.~\ref{fig:pbsmmx} we
show the solar modulated fluxes versus the neutralino mass.  We see
the same general trend as for the interstellar fluxes,
Fig.~\ref{fig:pbismx}, but we also see that there are many models with
fluxes above the {\sc Bess} measurements. However, this conclusion
depends strongly on which range one allows for the neutralino relic
density. In Fig.~\ref{fig:pbsmmx} we have coded the symbols according
to the relic density interval. As can be seen, essentially all models
which are in the {\sc Bess} measurement band have a relic density
$\Omega_\chi h^2 < 0.1$.  If we instead require $0.1 \lsim
\Omega_{\chi} h^2 \lsim 0.2$ the rates are never higher than the
measured flux.

This points to a weakness of this indirect method of detecting
supersymmetric dark matter: once the predicted rate is lower than the
presently measured flux, the sensitivity to an exotic component is
lost. This is because of the lack of a distinct signature which could
differentiate between the signal and the background. Alternative
indirect search methods, like gamma rays from the halo (see e.g.\
\cite{bub}), or neutrinos from the Sun or the Earth (see e.g.\
\cite{beg}) have the added virtue of giving both a directional and a
spectral signature which can be used to improve the signal to
background ratio well beyond the limits of present-day measurements.

The highest values for the fluxes in Fig.~\ref{fig:pbsmmx} (a) are 4
times higher than the {\sc Bess} measurement. However, the uncertainty
coming from the local halo density alone is larger than this. Given
the total mass of the galaxy, and restricting to our choice of halo
profile (isothermal sphere with $a=3.5$~kpc), we find a minimal local
halo density of 0.14 GeV/cm$^3$ which would correspond to a flux
reduction of a factor of 4.6. To that one should add the uncertainties
of the Monte Carlo simulations (up to a factor of 2) and the halo
profile, the propagation model and solar modulation.  For this reason
it is presently not possible to exclude any supersymmetric model on
the basis of antiproton measurements alone.

\subsubsection{Example of models}

In Table~\ref{tab:examples} we show 7 MSSM models that all give high
$\bar{p}$ fluxes. These models have acceptable relic densities, cover
a large mass range and have varying composition (and obey present
accelerator bounds).

\begin{figure}[tb]
\centerline{\epsfig{file=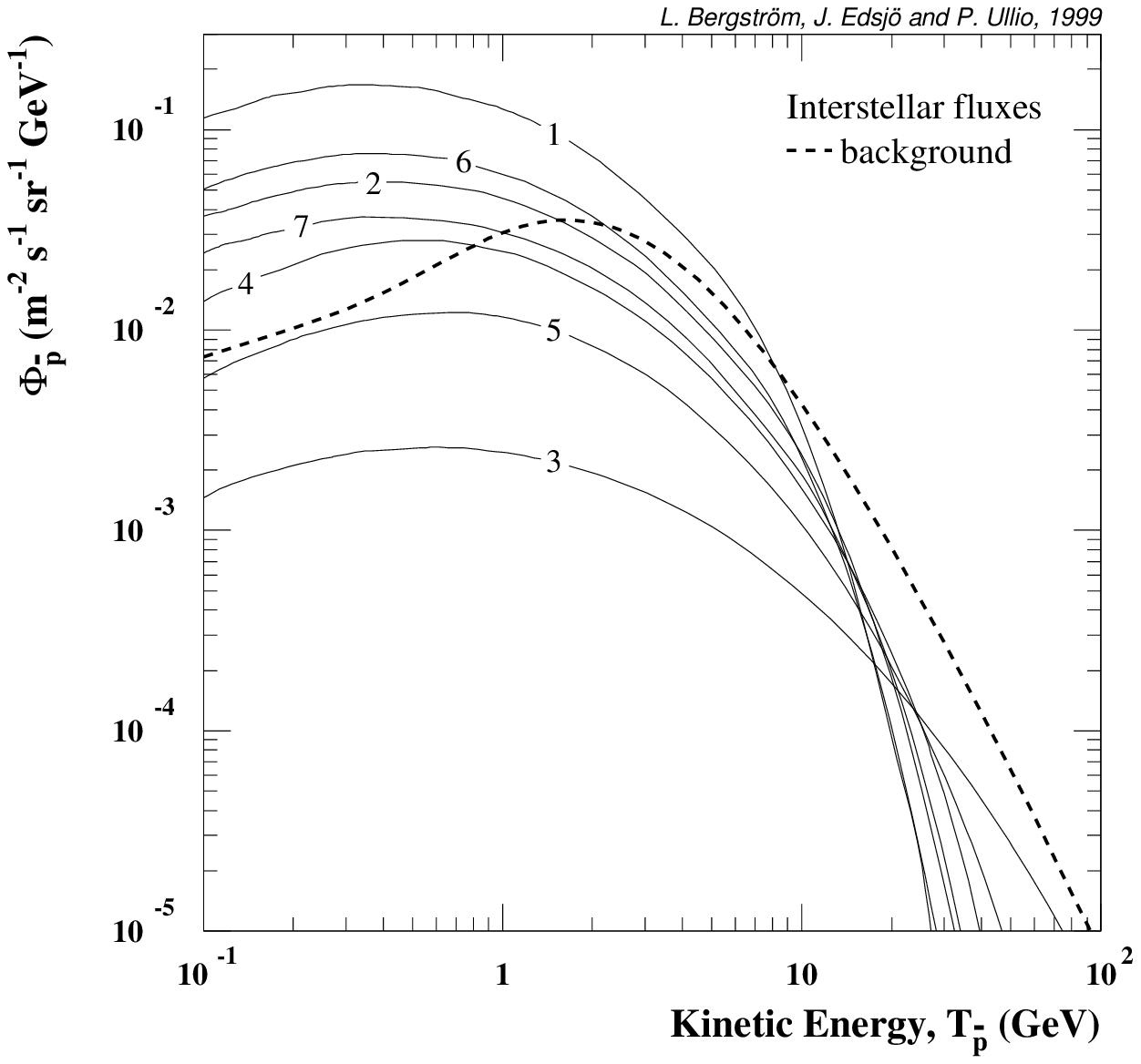,width=0.49\textwidth}
\epsfig{file=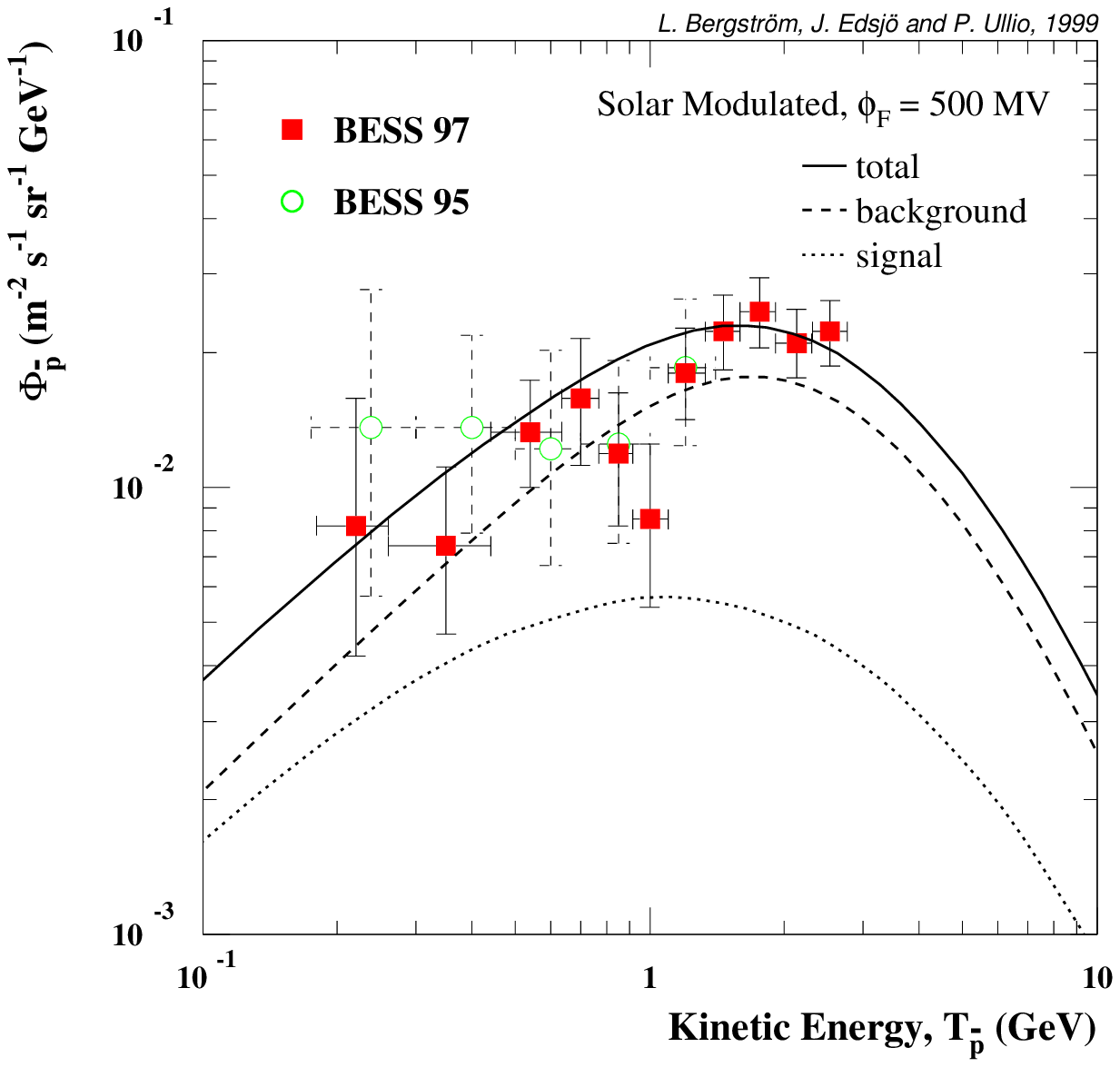,width=0.49\textwidth}}
\caption{
(a) Antiproton spectra for all 7 models appearing in
Table~\ref{tab:examples}.
(b) Example of a composite spectrum consisting of our reference
background $\bar p$ flux (Fig.\,\ref{fig:back}) reduced by 24 \% with
the addition of the predicted flux from annihilating dark matter neutralinos
of MSSM model number 5 in Table~\ref{tab:examples}.
}
\label{fig:nicemodel}
\end{figure}

In Fig.~\ref{fig:nicemodel} (a) the predicted differential $\bar p$ 
flux is shown for the 7 models.  They show maxima occurring at lower 
energies than for our canonical background.  At higher energies, the 
trend is that the slope of the flux decreases as the neutralino mass 
increases.  Model number 3 corresponds to a heavy neutralino and its 
spectrum is significantly less steep than the background.  If such a 
spectrum is enhanced, for instance by changing the dark matter density 
distribution, we would get a bump in the spectrum above 10 GeV\@.

In Table~\ref{tab:examples} we also show the annihilation rate and
the most important branching ratios. With the help of the results in
the earlier Sections, the antiproton flux from neutralino annihilation
can be derived. The difference between the parameterizations given in
Section \ref{sec:sim} and the full simulation results (that we have
used) is typically less than 20\%.

Also note that the branching ratio to $gg$ is never important for our
high-rate models (not only the ones in the table, but all high-rate
models). This is not in agreement with the results found by Jungman \&
Kamionkowski (1994). The reason for this difference is that we use the
improved $gg$ annihilation cross section of Bergstr\"om and Ullio
(1997).

The table also contains an indication of the rates for other detection
methods. The neutrino-induced muon flux in neutrino telescopes does
not show a strong correlation with the $\bar{p}$ flux, and it is
possible to find models that give either low or high rates.  Current
limits are about $10^3$--$10^4$ muons km$^{-2}$ yr$^{-1}$. We also
give the spin-indepedent neutralino-nucleon cross section (\cite{bg}),
which should be compared with the current limits that are of the order
of $10^{-5}$ pb. These show a better, but not perfect, correlation
with the $\bar{p}$ fluxes. The correlation is even stronger between
the $\bar{p}$ flux and both the $e^+$ flux and the $\gamma$ flux with
continuum energy spectrum. Both of these do not decrease as much with
neutralino mass as the antiproton flux does, however. For more details,
see Baltz \& Edsj{\"o} (1999) and \cite{clumpy}.  The cross section
for annihilation into monochromatic $\gamma$s (through $\gamma\gamma$
and $Z\gamma$) are uncorrelated with the $\bar{p}$ flux.

Finally, in Fig.~\ref{fig:nicemodel} (b) we show an example of a
hypothetical composite spectrum which consists of our canonical
background flux decreased by 24 \% (obtained e.g.\ by decreasing the
primary proton flux by $1\sigma$), and the signal for model 5 in
Table~\ref{tab:examples}. We can obtain a nice fit to the {\sc Bess}
data, but as noted before, there are no special features in the
spectrum that allow us to distinguish between this case and the case
of no signal.

\section{Discussion and conclusions}

We have seen that there is room, but no need, for a signal in the
measured antiproton fluxes. We have also seen that the optimal energy
to look for when searching for antiprotons is either below the solar
modulation cut-off or at higher energies than currently
measured. However, there are no special spectral features in the
signal spectra compared to the background, unless the signal is
enhanced and one looks at higher energies (above 10 GeV)\@.

We have stressed the somewhat disappointing fact that since the
present measurements by the {\sc Bess} collaboration already exclude a much
higher $\bar p$ flux at low energies than what is predicted through
standard cosmic-ray production processes, an exotic signal could be
drowned in this background. Even if it is not, the similar shape of
signal and background spectra will make it extremely hard to claim an
exotic detection even with a precision measurement, given the large
uncertainties in the predicted background flux (at least a factor of a
few, up to ten in a conservative approach).  We note that some of the
uncertainties may be reduced if an extrasolar probe aimed at
low-energy detection would be launched, a possibility that has been
recently proposed (\cite{spaceprobe}).

Although it is tempting to conclude that what has been measured by the
{\sc Bess} experiment is the standard cosmic-ray induced background
flux of antiprotons, one should keep in mind that it could, on the
contrary, be almost entirely due to an exotic source like neutralino
annihilation.  Since this possibility cannot be excluded (at least
until the problem of the dark matter in the Galactic halo has been
solved), one has to be cautious about using the measured antiproton
flux to deduce properties of antiproton propagation and, as has
recently been done (Geer \& Kennedy 1998), the antiproton lifetime. We
have checked that, using one of our high-mass neutralino models and a
clumpy distribution of dark matter in the halo, we can get an
excellent fit to the {\sc Bess} data for antiproton lifetimes as low
as $10^5$ years, clearly violating the claimed lower bound of Geer \&
Kennedy (1998). (For details, see Ullio (1999).)

\section*{Acknowledgements}

We thank Mirko Boezio, Alessandro Bottino and collaborators, Per Carlson and
Tom Gaisser for useful discussions, Paolo Gondolo for collaboration on
many of the numerical routines used in the supersymmetry part and
Markku J\"a\"askel\"ainen for discussions at an early stage of this
project.  L.B. was supported by the Swedish Natural Science Research
Council (NFR).  


\end{document}